\documentclass[twocolumn,apjl,times]{aastex62}

\usepackage[utf8]{inputenc}

\usepackage{ifthen}
\usepackage{etoolbox}
\usepackage{graphicx}
\usepackage{amsmath,amssymb}
\usepackage{color}
\usepackage{units}
\usepackage{hyperref}
\usepackage{xspace}

\newcommand{\macro}[1]{\textcolor{brown}{#1}\xspace}

\newcommand{\aligo}{Advanced LIGO (aLIGO)\renewcommand{\aligo}{aLIGO\xspace}\xspace}
\newcommand{\thebns}{\macro{GW170817}}
\newcommand{\gw}[1][]{gravitational wave#1 (GW#1)\renewcommand{\gw}[1][]{GW##1\xspace}\xspace}
\newcommand{\ns}[1][]{neutron star#1 (NS#1)\renewcommand{\ns}[1][]{NS##1\xspace}\xspace}
\newcommand{\bns}[1][]{binary neutron star#1 (BNS#1)\renewcommand{\bns}[1][]{BNS##1\xspace}\xspace}
\newcommand{\cw}[1][]{continuous wave#1 (CW#1)\renewcommand{\cw}[1][]{CW##1\xspace}\xspace}
\newcommand{\eos}[1][]{equation#1 of state (EoS)\renewcommand{\eos}[1][]{EoS\xspace}\xspace}
\newcommand{\elmag}[1][]{electromagnetic (EM)\renewcommand{\elmag}[1][]{EM\xspace}\xspace}
\newcommand{\otwo}{second observing run (O2)\renewcommand{\otwo}{O2\xspace}\xspace}
\newcommand{\sft}[1][]{Short Fourier Transform#1 (SFT#1)\renewcommand{\sft}[1][]{SFT##1\xspace}\xspace}
\newcommand{\snr}[1][]{signal-to-noise ratio#1 (SNR#1)\renewcommand{\snr}[1][]{SNR##1\xspace}\xspace}

\newcommand{\stamp}{STAMP\xspace}

\newcommand{\hmm}{HMM\xspace} % currently first called inside an enum, so the renew stuff doesn't work
\newcommand{\skyhough}{ATrHough\xspace}
\newcommand{\freqhough}{FreqHough\xspace}

\newcommand{\distanceGW}{\macro{\ensuremath{38_{-18}^{+8}}\,Mpc}} % free-sky-position TF2 high spin / low spin runs (both agree) released with GW170817PE paper (Fig3)
\newcommand{\distanceGWcompact}{\macro{40\,Mpc}}

\newcommand{\hostGalaxy}{\macro{NGC4993}}
\newcommand{\RunEndDate}{\macro{2017-08-26}} % Aug 26 2017 02:31:10 UTC
\newcommand{\RemainingRunLengthDays}{\macro{8.5}}

\newcommand{\Msun}{M_\odot}
\newcommand{\fgw}{f_\mathrm{gw}}
\newcommand{\fzero}{f_{\mathrm{gw}0}}
\newcommand{\fdotgw}{\dot{f}_{\mathrm{gw}}}
\newcommand{\fdotzero}{\dot{f}_{\mathrm{gw}0}}
\newcommand{\Tcoh}{\Tsft} % for simplicity of notation, equate this with \Tsft
\newcommand{\Tobs}{T_\mathrm{obs}}
\newcommand{\Izz}{I_{zz}}
\newcommand{\Egw}{E_\mathrm{gw}}
\newcommand{\Erot}{E_\mathrm{rot}}
\newcommand{\Etot}{E_\mathrm{tot}}
\newcommand{\Tsft}{T_\mathrm{SFT}}
\newcommand{\Tfft}{\Tsft} % for FreqHough, use same notation?
\newcommand{\Tdrift}{T_\mathrm{drift}}
\newcommand{\twait}{t_\mathrm{wait}}
\newcommand{\dUL}{d^{\ULperc}}
\newcommand{\EUL}{\Egw^{\ULperc}}
\newcommand{\hrss}{h_\mathrm{rss}}
\newcommand{\pfa}{p_{\mathrm{FA}}}
\newcommand{\tref}{t_\mathrm{start}}
\newcommand{\fref}{f_\mathrm{start}}
\newcommand{\fdotref}{\dot{f}_\mathrm{start}}
\newcommand{\tc}{t_\mathrm{c}}
\newcommand{\olddUL}{d^{50\%\xspace}} % 50% distance confidence level

\newcommand{\ULperc}{90\%\xspace} % upper limits confidence level
\newcommand{\bestULdistance}{\macro{\ensuremath{1\,\mathrm{Mpc}}}}
\newcommand{\bestULenergy}{\macro{\ensuremath{8\,\Msun c^2}}}

\newcommand{\TstartSTAMP}{\macro{1187008942}}
\newcommand{\TendSTAMP}{\macro{1187733618}}
\newcommand{\loudestHighFCandidateVLT}{\macro{3.07}}
\newcommand{\loudestLowFCandidateVLT}{\macro{3.18}}
\newcommand{\loudestHighFpvalVLT}{\macro{0.80}}
\newcommand{\loudestLowFpvalVLT}{\macro{0.81}}
\newcommand{\snratpvalthreshLowFVLT}{\macro{\gtrsim4.9}} % compare entry '4.9	0.050582' in results/VLT_lowf_bkg.txt
\newcommand{\snratpvalthreshHighFVLT}{\macro{\gtrsim3.5}} % compare entry '3.5	0.07166' in results/VLT_highf_bkg.txt

\newcommand{\TobsHMM}{\macro{9688\,s}}
\newcommand{\TstartHMM}{\macro{1187008882}}
\newcommand{\TendHMM}{\macro{1187018570}}
\newcommand{\NsftsHMM}{\macro{9688}}
\newcommand{\tauminHMM}{\macro{10^2}}
\newcommand{\taumaxHMM}{\macro{10^4}}
\newcommand{\loudestCandidateHMM}{\macro{2.6749}}
\newcommand{\lowestpvalHMM}{\macro{0.01}}

\newcommand{\TobsSkyHough}{\macro{1\,day}}
\newcommand{\TstartSkyHough}{\macro{1187008882}}
\newcommand{\TendSkyHough}{\macro{1187095282}}

\newcommand{\NcandsSkyHough}{\macro{51}}

\newcommand{\NcandsFreqHough}{\macro{521}}
\newcommand{\toffset}{\Delta t}

\newcommand{\tcGPSint}{\macro{1187008882}}
\newcommand{\tcGPSdecimal}{\macro{1187008882.443}}
\newcommand{\supraMaxLife}{\macro{\ensuremath{10^4}\,s}}
\newcommand{\Izzfiducial}{\macro{\ensuremath{4.34\times10^{38}}\,\mathrm{kg}\,\mathrm{m}^2}} % 100 in units of G=c=M_sol=1 => conversion factor 4.337e+36
\newcommand{\EtotValue}{\macro{\ensuremath{3.265\Msun c^2}}}

\newcommand{\dutycycleHday}{\macro{70\%}}
\newcommand{\dutycycleHfull}{\macro{83\%}}
\newcommand{\dutycycleHLday}{\macro{62\%}}
\newcommand{\dutycycleHLfull}{\macro{75\%}}
\newcommand{\dutycycleLday}{\macro{78\%}}
\newcommand{\dutycycleLfull}{\macro{85\%}}

\renewcommand{\macro}[1]{#1\xspace} % suppress highlight color for macros
\newcommand{\dcc}{LIGO-P1800195\xspace}
\journalinfo{} % suppress 'Draft version' in first page top-left header

\setcitestyle{notesep={}}

\newtoggle{fullauthorlist}
\toggletrue{fullauthorlist}
\newtoggle{endauthorlist}
\toggletrue{endauthorlist}
\newtoggle{fulltables}
\togglefalse{fulltables}

\begin{document}

\title{Search for gravitational waves from a long-lived remnant \\
       of the binary neutron star merger GW170817}
% \date{\today, revision \input{../gitID.txt}} % if you're compiling manually and LaTeX fails here, please run 'make' at least once in this directory!
\date{2019-04-11; report no. \dcc} % for arXiv/journal submission
% \linenumbers

\iftoggle{endauthorlist}{
  %
  % Put the author list at the end of the document.
  % Save author, affiliation, and maketitle commands.
  %
  \let\mymaketitle\maketitle
  \let\myauthor\author
  \let\myaffiliation\affiliation
  \author{The LIGO Scientific Collaboration}
  \author{The Virgo Collaboration}
}{
  %
  % Keep the author list on the initial title page.
  %
  \iftoggle{fullauthorlist}{
   %\input{LSC-Virgo-Authors-Feb-2018-aas.tex}
   %% LSC authorlist in AAS format
% \documentclass[modern]{aastex61}
\AuthorCollaborationLimit=3000  % AAS default is too small!
% \begin{document}

% \title{LSC February 2018 and Virgo February 2018 author list---LIGO-M1800028\\
% 5/28/2018. AAS style}

\author{B.~P.~Abbott}
\affiliation{LIGO, California Institute of Technology, Pasadena, CA 91125, USA}
\author{R.~Abbott}
\affiliation{LIGO, California Institute of Technology, Pasadena, CA 91125, USA}
\author{T.~D.~Abbott}
\affiliation{Louisiana State University, Baton Rouge, LA 70803, USA}
\author{F.~Acernese}
\affiliation{Universit\`a di Salerno, Fisciano, I-84084 Salerno, Italy}
\affiliation{INFN, Sezione di Napoli, Complesso Universitario di Monte S.Angelo, I-80126 Napoli, Italy}
\author{K.~Ackley}
\affiliation{OzGrav, School of Physics \& Astronomy, Monash University, Clayton 3800, Victoria, Australia}
\author{C.~Adams}
\affiliation{LIGO Livingston Observatory, Livingston, LA 70754, USA}
\author{T.~Adams}
\affiliation{Laboratoire d'Annecy de Physique des Particules (LAPP), Univ. Grenoble Alpes, Universit\'e Savoie Mont Blanc, CNRS/IN2P3, F-74941 Annecy, France}
\author{P.~Addesso}
\affiliation{University of Sannio at Benevento, I-82100 Benevento, Italy and INFN, Sezione di Napoli, I-80100 Napoli, Italy}
\author{R.~X.~Adhikari}
\affiliation{LIGO, California Institute of Technology, Pasadena, CA 91125, USA}
\author{V.~B.~Adya}
\affiliation{Max Planck Institute for Gravitational Physics (Albert Einstein Institute), D-30167 Hannover, Germany}
\affiliation{Leibniz Universit\"at Hannover, D-30167 Hannover, Germany}
% ...

\collaboration{The LIGO Scientific Collaboration and the Virgo Collaboration}

% \begin{abstract}
% This is the LSC February 2018 and Virgo February 2018 author list---LIGO-M1800028.
% \end{abstract}

% \section{Introduction}
% AAstex needs the $\backslash$section command and a few words here, or it will not produce output....

% \end{document}
 % this version has opt-ins/outs
  }{
    \author{The LIGO Scientific Collaboration}
    \author{The Virgo Collaboration}
  }
}

\begin{abstract}
One unanswered question about the binary neutron star coalescence \thebns is the nature of its post-merger remnant.
A previous search for post-merger gravitational waves targeted high-frequency signals from a possible neutron star remnant
with a maximum signal duration of 500\,s.
Here we revisit the neutron star remnant scenario
with a focus on longer signal durations
up until the end of the second Advanced LIGO-Virgo observing run,
\RemainingRunLengthDays\,days after the coalescence of \thebns.
The main physical scenario for such emission
is the power-law spindown of a massive magnetar-like remnant.
We use four independent search algorithms
with varying degrees of restrictiveness on the signal waveform
and different ways of dealing with noise artefacts.
In agreement with theoretical estimates,
we find no significant signal candidates.
Through simulated signals, we quantify
that with the current detector sensitivity,
nowhere in the studied parameter space
are we sensitive to a signal from more than \bestULdistance away,
compared to the actual distance of \distanceGWcompact.
This study however serves as a prototype for post-merger analyses
in future observing runs with expected higher sensitivity.

\end{abstract}


\section{Introduction}
\label{sec:intro}

The \bns observation \thebns~\citep{Abbott:2017oio} was
the first multimessenger astronomy event jointly detected in \gw[s]
and at many \elmag wavelengths~\citep{GBM:2017lvd}.
It originated remarkably close to Earth,
with a distance of \distanceGW
\footnote{Updated distance estimate
corresponding to Fig. 3 of \citet{Abbott:2018wiz},
where the sky location of the counterpart is not assumed,
hence differing slightly from the one quoted in the text for fixed-location runs.}
as measured by the LIGO and Virgo \gw detectors~\citep{TheLIGOScientific:2014jea,TheVirgo:2014hva} alone
and consistent \elmag distance estimates for
the host galaxy \hostGalaxy~\citep{Sakai:1999aw,Freedman:2000cf,Hjorth:2017yza,Lee:2018gfe}.

A \bns merger is expected to leave behind a remnant compact object,
either a light stellar-mass black hole or a heavy \ns,
which can emit a variety of post-merger \gw signals.
These are more difficult to detect than the pre-merger inspiral signal,
but the nearby origin of \thebns has still generated interest in searching for a post-merger signal.
Identifying the nature of the remnant
would be highly valuable for improving, among other things, constraints on
the nuclear \eos~\citep{Margalit:2017dij,Bauswein:2017vtn,Rezzolla:2017aly,Radice:2017lry}
over those obtained from the inspiral alone~\citep[e.g.,][]{Abbott:2017oio,Abbott:2018exr,Abbott:2018wiz}.

\citet{Abbott:2017dke} presented
a first model-agnostic search for
short ($\lesssim 1$\,s)
and intermediate-duration ($\lesssim500$\,s)
\gw signals.
No signal candidates were found.
The search sensitivity was estimated for several \gw emission mechanisms:
oscillation modes of a short-lived hypermassive \ns,
bar-mode instabilities,
and rapid spindown powered by magnetic-field induced ellipticities.
For all mechanisms,
a realistic signal from a \ns remnant of \thebns could only have been detected
with at least an order of magnitude increase in detector strain sensitivity.
A seconds-long postmerger signal candidate was reported by~\citet{vanPutten:2018abw}
with an estimated \gw energy lower than the sensitivity estimates of~\citet{Abbott:2017dke}.

An additional analysis in~\citet{Abbott:2018wiz}
used a Bayesian wavelet-based method
to put upper limits on the energy and strain spectral densities
over 1\,s of data around the coalescence.
These strain upper limits
are 3--10 times above the numerical relativity expectations
for post-merger emission from a hypermassive \ns at \distanceGWcompact.

In this paper, we focus on a long-lived \ns remnant,
covering possible signal durations which at the long end
are limited by the end of the \otwo on \RunEndDate,
giving a total data set spanning \RemainingRunLengthDays days from merger.
The shortest signal durations we cover
are $\sim$~hundreds of seconds after merger,
so that the new search presented here only partially overlaps
with the intermediate-duration search from~\citet{Abbott:2017dke}.
We assume the sky location of the \elmag counterpart~\citep{Coulter:2017wya,GBM:2017lvd}.

From considerations of realistic remnant \ns properties,
detailed in Sec.~\ref{sec:physics},
we do not expect to make a detection with this search.
Instead, the goal---as before in~\citet{Abbott:2017dke}---is
mainly to make sure that no unexpected signal is missed
in the longer-duration part of the parameter space.
This study also serves as a rehearsal for future post-merger searches
with improved detectors.
Hence, we use four search methods
with varying restrictiveness on the signal shape
and different ways of dealing with noise artefacts:
two generic unmodeled algorithms
and two that use templates based on
a power-law spin-down waveform model.

The \emph{Stochastic Transient Analysis Multidetector Pipeline} \citep[\stamp,][]{Thrane:2010ri}
is an unmodeled method using cross-power spectrograms. It
was already used for the intermediate-duration analysis in~\citet{Abbott:2017dke},
but is employed here in a different configuration optimized for much longer signal lengths.

The other three algorithms are derived from methods originally developed to search
for \cw[s]: persistent, nearly-monochromatic \gw signals
from older \ns[s]. \citep[For reviews, see][.]{Prix:2009oha,Riles:2017evm}.
Some \cw searches have targeted
relatively young \ns[s]~\citep{Aasi:2014ksa,Sun:2016idj,Zhu:2016ghk},
and adaptations of CW search methods to long-duration transient signals have been suggested
before~\citep{Prix:2011qv,Keitel:2015ova}.
However, the present search is the first time
that any \cw algorithms
have been modified in practice (on real data)
to deal with transients of rapid frequency evolution.

Specifically, these three are
\emph{Hidden Markov Model} (\hmm) tracking~\citep{Suvorova:2016rdc,Sun:2017zge}---a
template-free algorithm previously used to search for \cw[s] from the binary Scorpius X-1~\citep{Abbott:2017hbu}---and
two new model-dependent methods---\emph{Adaptive Transient Hough}~\citep[\skyhough,][]{Oliver:2019ksl}
and \emph{Generalized FrequencyHough}~\citep[\freqhough,][]{Miller:2018genfreqhough}---based on
algorithms~\citep{Krishnan:2004sv,T070124,Aasi:2013lva,Palomba:2005fp,Antonucci:2008jp,Astone:2014esa}
previously used in \cw all-sky searches~\citep[e.g., most recently in][]{Abbott:2017mnu,Abbott:2018bwn}.

After
discussing the astrophysical motivation and context for this search in Sec.~\ref{sec:physics},
presenting the analyzed data set in Sec.~\ref{sec:data}
and the four search methods in Sec.~\ref{sec:methods},
we discuss the combined search results in Sec.~\ref{sec:results}
and conclude with remarks on future applications in Sec.~\ref{sec:conclusion}.
Additional results
and details on the search methods
are given in the appendices.

\section{Astrophysical background and waveform model}
\label{sec:physics}

The probability for a long-lived \ns remnant
after a \bns merger
depends
on the progenitor properties and
on the nuclear \eos~\citep{Baiotti:2016qnr,Piro:2017zec}.
Using the progenitor masses and spins as measured from the inspiral~\citep{Abbott:2017oio,Abbott:2018wiz},
for many \eos the preferred scenarios are prompt collapse to a black hole
or the formation of a hypermassive \ns whose mass cannot be supported by uniform rotation
and thus collapses in $\lesssim1$\,s~\citep{Monitor:2017mdv}.
However, a supramassive \ns---less massive,
but above the maximum mass of a non-rotating \ns and stable for up to $\sim$\supraMaxLife~\citep{Ravi:2014}---or
even a long-time stable \ns
could also be consistent with some physically-motivated \eos
which allow for high maximum masses.

From the \elmag observational side,
circumstantial evidence
points towards a short-lived hypermassive
\ns~\citep{Kasen:2017sxr,Granot:2017tbr,Granot:2017gwa,Pooley:2017mzo,Matsumoto:2018mra};
though several authors~\citep{Yu:2017syg,Ai:2018jtv,Geng:2018vaa,Li:2018hzy}
consider continued energy injection from a long-lived remnant \ns.
Given this inconclusive observational situation,
we agnostically consider the possibility of \gw emission
from a long-lived remnant \ns
and seek here to constrain it from the LIGO data.

In two of our search methods,
and to estimate search sensitivities with simulations,
we use a waveform model~\citep{T1700408,Sarin:2018vsi}
originating from the general torque equation for the spindown of a rotating \ns:
\begin{equation}
 \label{eq:torque}
 \dot{\Omega} = -k \, \Omega^{n}.
\end{equation}
Here, \mbox{$\Omega= 2\pi f$} and $\dot{\Omega}$ are the star's angular frequency
and its time derivative, respectively,
and $n$ is the braking index.
A value of \mbox{$n\leq3$} corresponds to spindown predominantly through magnetic dipole radiation
and \mbox{$n=5$} to pure \gw emission~\citep{Shapiro1983}.
A braking index of \mbox{$n=7$} is conventionally associated with spindown through unstable $r$-modes~\citep[e.g.,][]{Owen:1998xg},
although the true value can be less for different saturation mechanisms~\citep{Alford:2014pxa,Alford:2012yn}.
The value of $k$ also depends on these mechanisms;
together with the starting frequency $\Omega_0$
it defines a spin-down timescale parameter
\begin{equation}
 \label{eq:k-tau}
 \tau = -\frac{\Omega_0^{1-n}}{k(1-n)} \,.
\end{equation}

Integrating Eq. (\ref{eq:torque})
and solving for the \gw frequency gives the \gw frequency evolution
\begin{equation}
 \label{eq:fgw}
 \fgw(t) = \fzero \left(1+ \frac{t}{\tau}\right)^{1/(1-n)},
\end{equation}
where \mbox{$\fgw=2f$},
$\fzero$ is the initial frequency
at a starting time $\tref$
(e.g., coalescence time $\tc$ of the \bns merger),
and $t$ is measured relative to $\tref$.

The dimensionless \gw strain amplitude
for a non-axisymmetric rotating body
following Eq.~\ref{eq:fgw}
is given by
\begin{equation}
 \label{eq:h0}
 h_0(t) = \frac{4\pi^2 G \Izz}{c^4} \frac{\epsilon}{d} \, \fzero^2 \left(1+ \frac{t}{\tau}\right)^{2/(1-n)}.
\end{equation}
Here, $\Izz$ is the principal moment of inertia,
$\epsilon$ is the ellipticity of the rotating body,
$d$ is the distance to the source,
$G$ is the gravitational constant,
and $c$ is the speed of light.
This model assumes that $n$, $\epsilon$ and $\Izz$ are constant
throughout the spin-down phase,
while in reality the spindown could be e.g. \gw-dominated at early times
and then transition into \elmag dominance,
and $\Izz$ can decrease with $\Omega$.

Our set of pipelines also allows for the power-law spindown model
to be valid for only part of the observation time:
To accomodate the possibility that
the newborn \ns has not immediately settled into a state that obeys the power-law model,
the \freqhough analysis starts
a few hours after the merger
(at \mbox{$\tc\approx\tcGPSdecimal$} in GPS seconds,
with the offset $\toffset$ varying across parameter space as described later),
making no assumption about the earlier NS evolution.
This provides complementary constraints to the other analyses.
The unmodeled \stamp search is also sensitive to signals
starting at either $\tc$ or at any later time,
as it does not impose a fixed starting time for any time-frequency tracks.
Moreover, neither \stamp nor \hmm impose the specific waveform model for their initial candidate selection.

The theoretical detectability of newborn NSs
evolving according to the spin-down model, Eq.~(\ref{eq:fgw}),
has been explored previously,
beginning with simple matched-filter estimates \citep{Palomba2001:mag,DallOsso:2008kll}.
More recent estimates also consider the limitations of practical searches
in the context of magnetars born following core-collapse supernovae \citep{DallOsso:2018dos}
and long-lived post-merger remnants~\citep{DallOsso:2014hpa,DallOsso:2018dos,Sarin:2018vsi},
finding qualitatively similar results.
With \aligo at design sensitivity~\citep{Aasi:2013wya}
and an optimal matched-filter analysis,
at \mbox{$d=$\,\distanceGWcompact}
an ellipticity \mbox{$\epsilon\sim10^{-2}$}
and timescale \mbox{$\tau\gtrsim10^{4}$\,s} would be required.
However, such large $\epsilon$ and long $\tau$
would imply more energy emitted than is available
from the remnant's initial rotation.
Considering actual data analysis pipelines applied to real detector data (at O2 sensitivity),
a detectable signal only seems possible
for extremely large $\epsilon \geq 0.1$ and short $\tau$
% due to constraints on the energy budget of the system
due to the energy budget constraint
\citep{Sarin:2018vsi}.
Such ellipticities are physically unlikely~\citep{JohnsonMcDaniel:2012wg}
and would require internal magnetic fields greater than $\sim10^{17}$\,G~\citep[e.g.,][]{Cutler:2002},
which might be intrinsically unstable~\citep{Reisenegger:2008yk}
and
for which very rapid EM-dominated spindown would be expected.

For r-modes, the \gw strain follows a different relation than Eq.~\ref{eq:h0}~\citep{Owen:1998xg},
but the physically relevant parameter (the saturation amplitude) is also expected to be small~\citep{Arras:2002dw,Bondarescu:2008qx}.
They could be an important emission channel especially at high frequencies,
and the search presented in this paper also covers braking indices up to $n=7$.
The sensitivity estimates presented in Sec.~\ref{sec:results} however,
for simplicity, will be for $n=5$ only.

Still, with this first search for long-duration post-merger signals,
we demonstrate that available analysis methods can comprehensively
cover the relevant parameter space,
and thus will be ready once detector sensitivity has improved,
or in the case of a fortunate, very nearby \bns event.

% \showthe\columnwidth % Use this to determine the width of a single-column figure.
% got: 242.26653pt
% \showthe\textwidth % Use this to determine the width of a double-column figure.
% got: 513.11743pt.
% \showthe\linewidth % equivalent to single or double column, dependent on format
% got: 242.26653pt

\section{Detectors and data set}
\label{sec:data}

In this analysis,
we use data from the two \aligo detectors
in Hanford, Washington (H1)
and Livingston, Louisiana (L1).
No data from Virgo~\citep{TheVirgo:2014hva} or GEO600~\citep{Dooley:2015fpa}
was used because of their lower sensitivity.{\footnote{E.g. at 500\,Hz
the noise strain amplitude spectral density was about a factor of $\sim5$ for Virgo
and $\sim20$ for GEO600 worse than for L1 in late O2.}
Three of the pipelines use data up to 2\,kHz;
\stamp also uses data up to 4\,kHz.
Both detectors in their O2 configuration had their best sensitivity
in the 100--200\,Hz range,
with significantly less sensitivity in the kHz range
(e.g. a factor $\sim$4 worse in strain at 2\,kHz)---see Fig.~\ref{fig:segments}.
For the lower analysis cutoff of each pipeline, see Sec.~\ref{sec:methods}.

\begin{figure}[t]
 \includegraphics{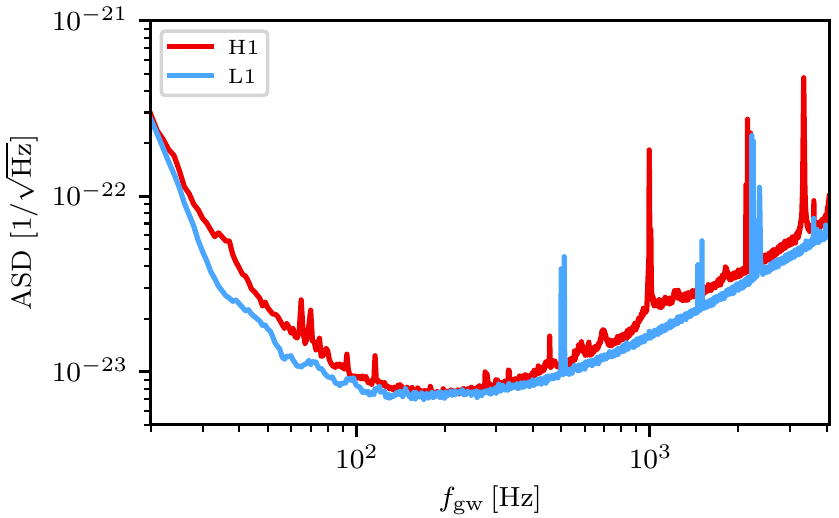}
 \includegraphics{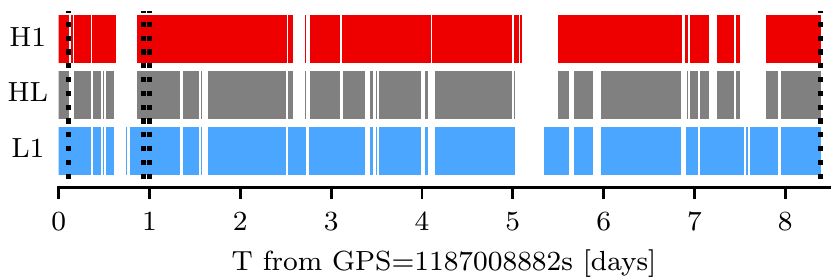}
 \caption{
  \label{fig:segments}
  Top panel: Noise strain amplitude spectral density (ASD) curves of LIGO Hanford (H1) and Livingston (L1)
  on August 17, 2018.
  (Averaged over 1800\,s stretches including \thebns.)
  Lower panel: Analysable science mode data segments
  for the remaining \otwo run
  after the \thebns event.
  Vertical dotted lines
  mark the analysis end times,
  from left to right,
  for \hmm, \freqhough, \skyhough and \stamp.
 }
\end{figure}

Starting from a rounded \thebns coalescence time of
$\tc\approx\tcGPSint$\,s,
the \hmm pipeline uses \TobsHMM of data
(until the first gap in H1 data),
\skyhough uses \TobsSkyHough of data after $\tc$,
\freqhough analyzes from 1 to about 18 hours after $\tc$ in different configurations,
and \stamp analyzes the whole \RemainingRunLengthDays\,days of data
until the end of \otwo on \RunEndDate.
The duty cycle of both detectors (H1, L1) was
100\% during the first \TobsHMM after the merger,
(\dutycycleHday,~\dutycycleLday) for the first day
(\dutycycleHLday in coincidence),
and (\dutycycleHfull,~\dutycycleLfull) for the full data set
(\dutycycleHLfull in coincidence).
The analysed data segments are also illustrated in Fig.~\ref{fig:segments}.
\stamp processes the $h(t)$ strain data into cross-power
time-frequency maps (see Sec.~\ref{sec:stamp} for details),
while for the other pipelines the basic analysis units
are \sft[s] of 1--8\,s duration.

Several known noise sources have been subtracted from the strain data
using a new automated procedure~\citep{Davis:2018yrz} applied to the full O2 data set,
processing a much larger amount of time than the cleaning method~\citep{Driggers:2018gii}
used for the shorter data sets analysed in previous \thebns publications.
Calibration uncertainties~\citep{Cahillane:2017vkb} for this data set
are estimated as below
4.3\% in amplitude and 2.3 degrees in phase
for 20--2000\,Hz,
and
4.5\% in amplitude and 3.8 degrees in phase
for 2--4\,kHz;
these are tighter than for the
initial calibration version used in \citet{Abbott:2017dke}.
These uncertainties are not explicitly propagated into
the sensitivity estimates presented in this paper,
since they are smaller than
other uncertainty contributions
and the degeneracies in amplitude parameters.

\section{Search methods and configurations}
\label{sec:methods}

Here we briefly describe the four search methods,
first the unmodeled \stamp and \hmm pipelines
and then the two Hough pipelines tailored to the power-law spindown model.
Additional details can be found in Appendix~\ref{sec:appendix-methods}.

\begin{deluxetable*}{lcccc}
 \tablecaption{
  \label{tbl:search_configs}
  Configurations of the four analysis pipelines used in this paper.
 }
 \tablehead{
  &
  \colhead{\stamp}     &
  \colhead{\hmm}       &
  \colhead{\skyhough}  &
  \colhead{\freqhough}
 }
 \startdata
  search start\tablenotemark{a}            & $\tc$                     & $\tc$              & $\tc$                        & $\tc+$ (1--7)\,hours\tablenotemark{b} \\
  search duration [hours]                  & 201.3\tablenotemark{c}    & 2.7                & 24                           & 2--18\tablenotemark{b} \\
  $\fgw$ data range [Hz]                   & 30--4000\tablenotemark{c} & 100--2000          & 187--2000                    & 50--2000 \\
  $n$ coverage                             & \textit{unmodeled}        & \textit{unmodeled} & 2.5--7.0                     & 2.5--7.0 \\
  $\fref$ coverage [Hz]\tablenotemark{d}   & \textit{unmodeled}        & \textit{unmodeled} & 500--2000                    & 500--2000 \\
  $\tau$ coverage [s]                      & \textit{unmodeled}        & \textit{unmodeled} & $10^2$--$10^5$               & $10$--$10^5$ \\
 \hline
 \emph{injection set for sensitivity estimation}\tablenotemark{e}      &                    &                              & \\
  signal start\tablenotemark{a}            & random                    & $\tc$              & $\tc$                        & $\tc+$ [1,2,5]\,hours\tablenotemark{b} \\
  $n$ coverage                             & 5.0                       & 2.5--7.0           & 5.0                          & 5.0 \\\
  $\fref$ coverage [Hz]\tablenotemark{d}   & 500--3000                 & 500--2000          & 550--2000                    & 390--2000 \\
  $\tau$ coverage [s]                      & $10^2$--$10^4$            & $10^2$--$10^4$     & $6\times10^2$--$3\times10^4$ & $4\times10^2$--$2\times10^4$ \\
  inclination $\cos\iota$                  & 0.0, 1.0                  & random             & 0.0, 1.0                     & random \\
 \enddata
 \tablenotetext{a}{Coalescence time $\tc\approx\tcGPSint$ rounded to integer GPS seconds.}
 \vspace{-0.25\baselineskip}
 \tablenotetext{b}{\freqhough search start and duration vary across parameter space.}
 \vspace{-0.25\baselineskip}
 \tablenotetext{c}{In separate maps of 15\,000\,s length
                   and 20--2000\,Hz and 2000--4000\,Hz configurations.}
 \vspace{-0.25\baselineskip}
 \tablenotetext{d}{$\fref=\fgw(t=0)$ for \hmm and \skyhough; $\fref=\fgw(t=\toffset)$ for \stamp and \freqhough.}
 \vspace{-0.25\baselineskip}
 \tablenotetext{e}{Discrete sets of injections within these ranges; not all combinations used.
                   See Sec.~\ref{sec:results} and the per-pipeline tables in the appendix for details.}
 \vspace{-\baselineskip}
\end{deluxetable*}

Each analysis uses the known sky location of the counterpart near
RA = 13.1634\,hrs, Dec. = $-23.3815^\circ$
\citep{Coulter:2017wya,GBM:2017lvd},
but makes different choices for the analysed data span.
The recovery efficiency of each algorithm
is studied with simulated signals under the waveform model
from Sec.~\ref{sec:physics},
as described in Sec.~\ref{sec:results} and Appendix~\ref{sec:appendix-pipeline-sens}.

A summary of configurations for all four pipelines,
both for the main search and the sensitivity estimation simulations,
is given in Table~\ref{tbl:search_configs}.

\subsection{STAMP}
\label{sec:stamp}

\stamp~\citep{Thrane:2010ri} is an unmodeled search pipeline designed to detect gravitational wave transients.
Its basic unit is a spectrogram made from cross-correlated data between two detectors.
Narrowband transient gravitational waves produce tracks of excess power within these spectrograms,
and can be detected by pattern recognition algorithms.
Each spectrogram pixel is normalized with the noise to obtain a \snr for each pixel.

\stamp was used in the first \thebns post-merger search~\citep{Abbott:2017dke}
in a configuration with 500\,s long spectrograms.
To increase sensitivity to longer \gw signals,
here we use spectrogram maps of 15\,000\,s length.
The search is split into two frequency bands from 30--2000\,Hz and 2000--4000\,Hz. 
The former uses pixels of $100\,\mathrm{s} \times 1\,\mathrm{Hz}$,
while the latter uses shorter-duration pixels of $50\,\mathrm{s} \times 1\,\mathrm{Hz}$
to limit \snr loss due to the Earth's rotation changing the phase difference between detectors.

We then use Stochtrack~\citep{Thrane:2013bea},
a seedless clustering algorithm,
to identify significant clusters of pixels within these maps. 
The algorithm uses one million quadratic B{\'e}zier curves as templates for each map,
and the loudest cluster is picked for each map.
More details about the pixel size choice, the detection statistic and the search results are in Appendix~\ref{sec:appendix-methods-stamp}.

The on-source data window is
from just after the time of the merger
to the end of \otwo
(\TstartSTAMP--\TendSTAMP).
To measure the background and estimate the significance of the clusters found,
we run the algorithm on time-shifted data
from June 24th to just before the merger.

\subsection{HMM tracking}
\label{sec:hmm}

Hidden Markov model (HMM) tracking provides a computationally efficient strategy
for detecting and estimating a quasimonochromatic \gw signal
with unknown frequency evolution and stochastic timing noise~\citep{Suvorova:2016rdc,Sun:2017zge}.
It was applied to data from the first \aligo observing run
to search for \cw[s] from the
low-mass X-ray binary Scorpius X-1~\citep{Abbott:2017hbu}.
The revision of the algorithm in \citet{Sun:2017zge}
is also well suited to searching for a long-transient signal
from a \bns merger remnant,
if the spin-down time-scale is in the range
\mbox{$\tauminHMM\,{\rm s} \lesssim \tau \lesssim \taumaxHMM$\,s}.

A HMM is an automaton based on a Markov chain
(a stochastic process
transitioning between discrete states at discrete times),
composed of a hidden (unmeasurable) state variable
and a measurement variable.
A HMM is memoryless,
i.e., the hidden state at time $t_{n+1}$
only depends on the state at time $t_n$,
with a certain transition probability.
The most probable sequence of 
hidden states given the observations
is computed by the classic Viterbi algorithm \citep{Viterbi:1967}.
Details on the probabilistic model
can be found in Appendix~\ref{sec:appendix-methods-hmm}.

In this analysis, we track the \gw signal frequency
as the hidden variable,
with its discrete states mapped one-to-one
to the frequency bins in the output of a frequency-domain estimator
computed over an interval of length $\Tdrift$.
We aim at searching for signals with
\mbox{$\tauminHMM\,{\rm s} \lesssim \tau \lesssim \taumaxHMM$\,s},
such that the first time derivative $\dot{f}_{\rm gw}$ of the signal frequency $f_{\rm gw}$ 
satisfies \mbox{$\dot{f}_{\rm gw} \approx f_{\rm gw}/\tau \lesssim 1$\,Hz\,s$^{-1}$},
given \mbox{$\Tdrift = 1$\,s} and a frequency bin width of
\mbox{$\Delta f = 1$\,Hz}.
The motion of the Earth with respect to the solar system barycenter (SSB)
can be neglected during a $\Tdrift$ interval.
Hence we use a running-mean
normalized power
in \sft[s] with length
\mbox{$\Tsft = \Tdrift =1$\,s}
as the estimator
to calculate the HMM emission probability.

We analyze \NsftsHMM\,s of data (GPS times \TstartHMM--\TendHMM)
in a 100--2000\,Hz frequency band with multiple configurations optimized for different $\tau$.
We do not analyze longer data stretches because
(i) several intervals in the data after GPS time \TendHMM are not in analysable science mode,
and (ii) signals with
\mbox{$\tauminHMM\,{\rm s} \lesssim \tau \lesssim \taumaxHMM$\,s}
drop below the algorithm's sensitivity limit after $\sim \taumaxHMM$\,s;
observing longer merely accumulates noise without improving \snr.
The \NsftsHMM~SFTs are Hann-windowed.
The detection statistic $\mathcal{P}$ is defined in Eqn.~\ref{eqn:det_stat}. 
The methodology and analysis is fully described in \citet{Sun:2018owi}.

\subsection{Adaptive Transient Hough}
\label{sec:skyhough}

The Adaptive Transient Hough search method is described in detail in \citet{Oliver:2019ksl}.
It follows a semi-coherent strategy similar to the SkyHough~\citep{Krishnan:2004sv,T070124,Aasi:2013lva}
all-sky \cw searches,
but adapted to rapid-spindown transient signals.

We start from data
in the form of Hann-windowed SFTs with lengths of [1,2,4,6,8]\,s,
covering one day after merger
(GPS times \TstartSkyHough--\TendSkyHough).
These are digitized by setting a threshold of 1.6 on their normalized power,
as first derived by~\citet{Krishnan:2004sv},
replacing each SFT by a collection of zeros and ones called a peak-gram.
For each point in parameter space,
the Hough number count is the weighted sum of the peak-grams
across a template track accounting for Doppler shift and the spindown of the source.
The use of weights minimizes the influence of time-varying
detector antenna patterns and noise levels~\citep{T070124}.
For this post-merger search it also accounts for the amplitude modulation
related to the transient nature of the signal.

The search parameter space for the model from Sec.~\ref{sec:physics} covers
a band of \mbox{500--2000\,Hz} in starting frequencies $\fzero$,
braking indices of \mbox{$2.5\le n\le 7$}
and spindown timescales of \mbox{$10^2\le \tau\le 10^5$\,s}.
The search runs over 16042 subgroups,
each containing a range of
150\,Hz in $f_0$,
0.25 in $n$
and $1000$\,s in $\tau$.
Each subgroup is analyzed
with the longest possible SFTs
according to the criterion~\citep{Oliver:2019ksl}
\begin{equation}
 \label{eq:skyhough_tcoh}
 \Tcoh \leq \frac{\sqrt{(n-1) \tau }}{\sqrt{\fgw}} \,,
\end{equation}
and for each template the observation time is selected as
\mbox{$\Tobs=\min(4\,\tau,24\,\mathrm{hours})$}.
Over the whole template bank, the search uses data from 187--2000\,Hz.

Each template is ranked based on the deviation of its weighted number count
from the theoretical expectation for Gaussian noise (the critical ratio)
as described in appendix~\ref{sec:appendix-methods-skyhough}.
The detection threshold corresponds to a two-detector
$5\sigma$ false alarm probability for the entire template bank.
A per-detector critical ratio threshold was also set
to check the consistency of a signal between H1 and L1.

\subsection{Generalized FrequencyHough}
\label{sec:freqhough}

The FrequencyHough is a pattern-recognition technique
originally developed to search for \cw[s]
by mapping points in time-frequency space of the detector
to lines in frequency-spindown space~\citep{Antonucci:2008jp,Astone:2014esa}.
This only works if the signal frequency varies in time very slowly.
\citet{Miller:2018genfreqhough} have generalized the FrequencyHough
for postmerger signals, where we expect much higher spindowns.

The search starts at a time offset \mbox{$\toffset=\tref-\tc$}
after coalescence time $\tc$,
so that the waveform model is interpreted
with starting frequency \mbox{$\fref=\fgw(t=\toffset)$}
taking the place of $\fzero$ in Eq.~\ref{eq:fgw}.
In this way,
assuming that
the \ns has already spun down before $\tref$
following some arbitrary track,
we would be probing higher initial frequencies and spindowns
through a less challenging parameter space during the search window.
Furthermore, the source parameters $(n,\fref,\tau$)
are transformed to new coordinates
such that in the new space the behavior of the signal is linear.
See appendix~\ref{sec:appendix-methods-freqhough} for the transformation relations.

We search across the parameter space with a fine, nonuniform grid:
For each braking index $n$, we do a Hough transform and then
record the most significant candidates over the parameter range of the resulting map.
This is done separately on the data from each detector,
and then we check candidates for coincidence between detectors
according to their Euclidean distance in parameter space.

The search is run in three configurations using varying \mbox{$\Tfft=2,4,8$\,s},
covering different observing times,
starting $\toffset = 1$--$7$\,hours after merger.
It covers \mbox{$n=[2.5,7]$}, \mbox{$\fref=[500,2000]$\,Hz} and \mbox{$\tau=[10,10^5]$\,s},
analyzing detector data from 50 to 2000\,Hz.

Candidates are also ranked by critical ratio (deviation from the theoretical expectation for Gaussian noise)
in this analysis.
Most can be vetoed by the coincidence step
or by considering detector noise properties;
a follow-up procedure for surviving candidates is also described in appendix~\ref{sec:appendix-methods-freqhough}.

\section{Search results and sensitivity estimates}
\label{sec:results}

\subsection{Absence of significant candidates}
\label{sec:results-candidates}

The four search methods all either
found no significant candidates in the \aligo data after \thebns;
or those that were found, were clearly vetoed as instrumental artifacts.

For the unmodeled \stamp search,
the loudest triggers in the low- and high-frequency bands
have \snr[s] of \loudestLowFCandidateVLT and
\loudestHighFCandidateVLT respectively.
The time-shifted backgrounds only just start to drop off near these \snr[s],
so that they correspond to false-alarm probabilities $\pfa$ of \loudestLowFpvalVLT and \loudestHighFpvalVLT
which are completely consistent with noise.
For reference, \mbox{$\pfa=0.05$}
would have only been reached for \snr[s] of $\snratpvalthreshLowFVLT$ and $\snratpvalthreshHighFVLT$
for these low- and high-frequency background distributions, respectively.
(See Fig.~\ref{fig:STAMP_fg_bg} in appendix~\ref{sec:appendix-methods-stamp}.)

For \hmm, the loudest trigger has a detection statistic
$\mathcal{P}=\loudestCandidateHMM$
(as defined in Eq.~\ref{eqn:det_stat}),
corresponding to a false-alarm probability of \lowestpvalHMM,
right below the threshold set beforehand as
significant enough for further study.
The trigger is found with observing time \mbox{$\Tobs=200$\,s}
starting from \mbox{$t=\tc$}.
Monte-Carlo simulations show that for signals that this setup is sensitive to,
higher $\mathcal{P}$ should be obtained
with longer $\Tobs$.
Follow-up analysis of the trigger with
\mbox{$300\,{\rm s} \leq \Tobs \leq 1000\,{\rm s}$}
confirms that it does not follow this expectation;
hence it is discarded as spurious.

\skyhough found \NcandsSkyHough initial candidates
over the covered part of $(n,\fzero,\tau)$ parameter space.
All of these were excluded
with the follow-up procedure
described in appendix~\ref{sec:appendix-methods-skyhough}
as inconsistent
with the expected spindown model
and more likely to be caused by monochromatic detector artifacts (lines)
contaminating the search templates.

The \freqhough search returned \NcandsFreqHough candidates
over the covered part of $(n,\fref,\tau)$ parameter space.
We vetoed 10 of them because they were within frequency bands
contaminated with known noise lines~\citep{Covas:2018oik}.
510 of the remaining candidates had much higher ($>4$ times) critical ratios in H1 than in L1,
which is inconsistent with true astrophysical signals
when considering the relative sensitivities, duty factors and antenna patterns.
There was one remaining candidate, with a critical ratio of 5.21 in H1 and 4.88 in L1,
which was followed up and excluded with the procedure described in appendix~\ref{sec:appendix-methods-freqhough}.

\subsection{Sensitivity estimates with simulated signals}
\label{sec:results-sensitivities}

Starting from this non-detection result,
we use simulated signals according to Eq.~\ref{eq:fgw}
to quantify the sensitivity of each analysis
given the data set around the time of \thebns
and its known sky location.
The sets of injected parameters are different for each pipeline,
and there are also some differences in procedure:
\stamp performs injections on the same data as the main search but with a non-physical time shift between
the detectors \citep[as in][]{Abbott:2017dke,Abbott:2017muc};
\hmm injects signals into the original set of \sft[s] but with randomly permuted timestamps;
and the other two pipelines inject signals into exactly the same data as analysed in the main search.
\hmm and \skyhough perform all injections starting at merger time $\tc$,
with $\fzero$ in Eq.~\ref{eq:fgw} interpreted as the frequency at $\tc$,
while injections for \freqhough are done at \mbox{$\toffset=1,2$} or $5$\,hours after $\tc$,
chosen as representative starting times for each search configuration,
and $\fzero$ correspondingly set at \mbox{$\tc+\toffset$}.
Similarly, \stamp
treats $\fzero$ as the starting frequency of each injection,
which have $\toffset$ distributed through the whole search range,
yielding a time-averaged sensitivity.
In the following, we use $\fref$ to refer to any of these choices.

These differences in injection procedure,
and different choices of detection threshold,
mean that any comparison of the following results
does not correspond to a representative evaluation of general pipeline performance,
but is solely in the interest of estimating how much sensitivity
is missing for a \thebns-like post-merger detection
based on the specific configurations as used in the present search.

\begin{figure*}[t]
 \includegraphics{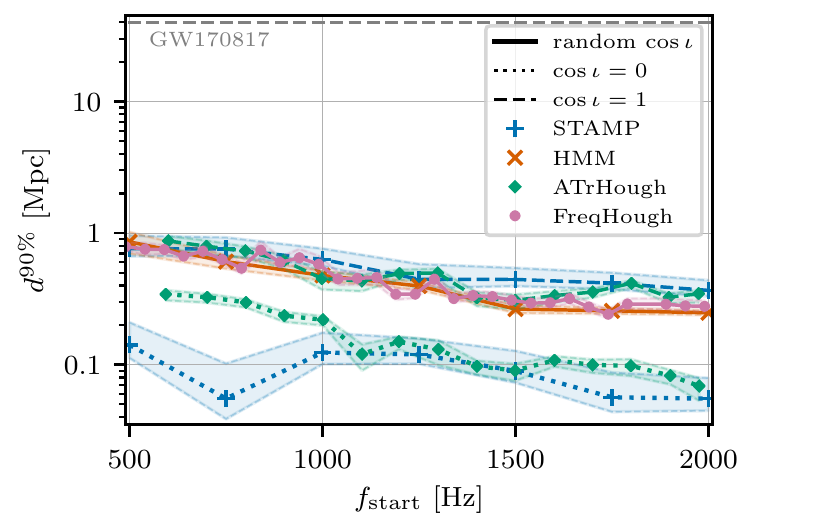}
 \includegraphics{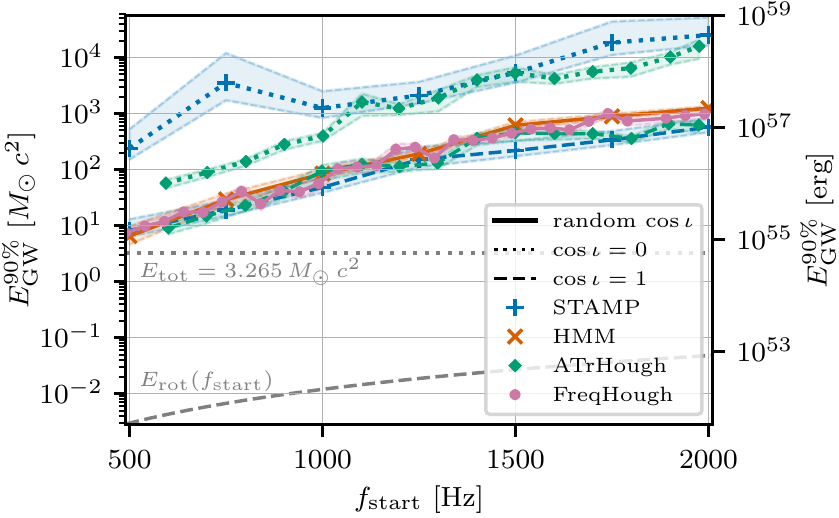}
 \caption{
  \label{fig:ULs_n5_dist_energy}
  A sample of search sensitivities achieved
  for the power-law spindown signal model
  with braking index \mbox{$n=5$}.
  Results are shown as sensitive distance $\dUL$
  (left panel)
  for otherwise physical parameters,
  or as required emitted energy $\EUL$
  at a fixed distance \mbox{$d=$\,\distanceGWcompact}
  (right panel),
  both as a function of reference starting frequency $\fref$
  used for the injections of each pipeline.
  \mbox{($\fref=\fgw(t=\toffset)$} for \stamp and \freqhough and
  \mbox{$\fref=\fzero=\fgw(t=0)$} for the others.) \\
  See Fig.~\ref{fig:UL_injs_coverage} for the parameter ranges covered
  by each injection set.
  This figure shows the subset with highest sensitivity for each analysis;
  this corresponds to the shortest (\mbox{$\tau=100$\,s}) injections
  for \stamp and \hmm,
  while for \skyhough and \freqhough
  \mbox{$\tau(\fref)$} is variable,
  depending on the search coherence length,
  as also listed in Tables \ref{tbl:ULs_n5_skyhough} and \ref{tbl:ULs_n5_freqhough}.
  Note that detection thresholds are also different between pipelines. \\
  The \ns ellipticity $\epsilon$ is always chosen as the maximum allowed by
  the energy budget constraint
  \mbox{$\Egw=\Erot$}
  at each ($n,\fref,\tau)$ parameter point,
  assuming a \ns moment of inertia of
  \mbox{$\Izz=100\,\Msun^3\,G^2/c^4\approx\Izzfiducial$}.
  Injections were randomized over source inclination $\cos\iota$ for \hmm and \freqhough,
  while for \stamp and \skyhough injections
  for the best case (\mbox{$\cos\iota=1$})
  and worst case (\mbox{$\cos\iota=0$})
  are shown separately. \\
  For comparison, the known distance to the source of \thebns
  is indicated by a horizontal dashed line in the left panel,
  as well as two (optimistic) energy upper limits in the right panel:
  the total system energy \citep[dotted line, using a fiducial value of $\Etot=\EtotValue$ as in][]{Abbott:2017dke}
  and the initial rotational energy $\Erot$ as a function of $\fref$ (dashed line).
 }
\end{figure*}

\begin{figure*}
 \includegraphics{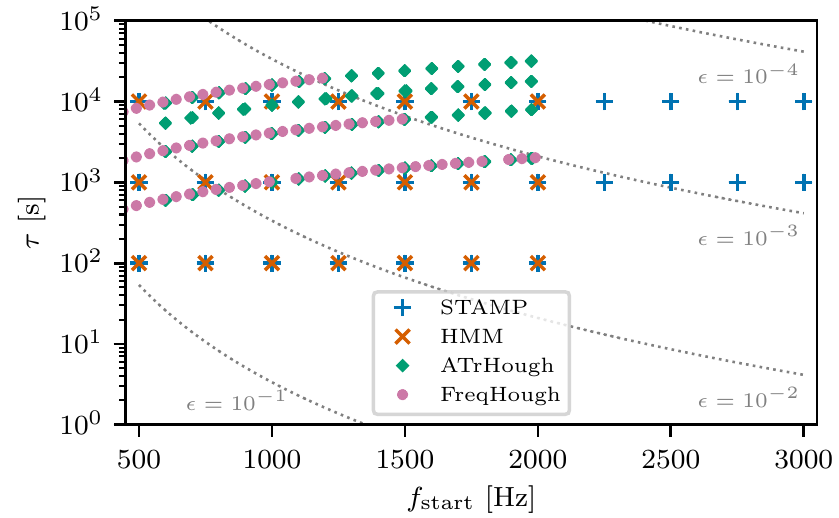}
 \includegraphics{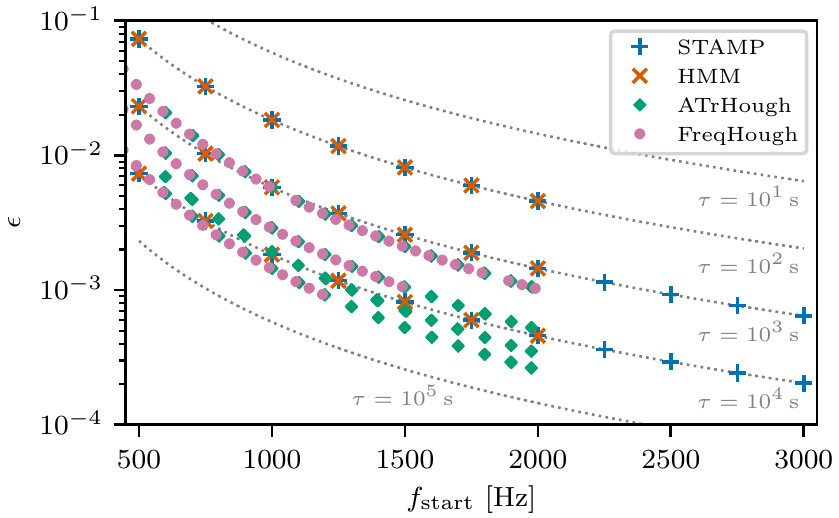}
 \caption{
  \label{fig:UL_injs_coverage}
  Parameter coverage in $\fref$, $\tau$ and $\epsilon$
  of the injection sets
  used for the \mbox{$n=5$} sensitivity estimates,
  as listed in Tables \ref{tbl:ULs_n5_stamp}--\ref{tbl:ULs_n5_freqhough}.
  As shown in the left panel,
  the \hmm and \stamp injections
  are at fixed \mbox{$\tau\in[10^2,10^3,10^4]$\,s},
  while for \skyhough and \freqhough
  different $\tau(\fref)$ curves are covered
  for different choices of $\Tsft$ (and, in the case of \freqhough,
  $\toffset$)
  in the search setup.
  At each $(n,\fref,\tau)$ parameter space point,
  the maximum $\epsilon$ allowed by the energy budget
  \mbox{($\Egw=\Erot$)}
  is chosen (right panel),
  assuming a \ns moment of inertia of
  \mbox{$\Izz=100\,\Msun^3\,G^2/c^4\approx\Izzfiducial$}.
  Lines of constant $\epsilon$ (left panel)
  or $\tau$ (right panel)
  are shown for comparison.
  \stamp injections include $\fref$ up to 3000\,Hz for longer $\tau$,
  with those above 2000\,Hz covered by the high-frequency search configuration.
  But for $\tau=100$\,s, we limit $\fref$ to 2000\,Hz because injections at higher
  frequencies would leave the high-frequency band too rapidly to be recoverable.}
\end{figure*}

We focus here on results for a braking index of \mbox{$n=5$},
as expected for spin-down dominated by \gw emission from a static quadrupole deformation.
The signal amplitude $h_0$ (as given in Eq.~\ref{eq:h0})
is degenerate between the ellipticity $\epsilon$,
moment of inertia $\Izz$ and distance $d$.
We choose a fiducial value of
\mbox{$\Izz=100\,\Msun^3\,G^2/c^4\approx\Izzfiducial$},
consistent with \eos yielding
a supramassive or stable remnant:
The high mass and assumed rapid rotation
can increase the moment of inertia by more than a factor of 3
compared to a nonrotating NS of $1.4 \, M_\odot$.
In addition, \eos compatible with the high remnant mass favor
larger moments of inertia already at lower mass.
For a given set of model parameters $\{n=5,\fref,\tau\}$
we consider the maximum $\epsilon$ allowed by the initial rotational energy budget~\citep{Sarin:2018vsi}:
the total emitted \gw energy as $t\rightarrow\infty$,
\begin{equation}
 \label{eq:Egw}
 \Egw = -\int\limits_{t=\tref}^{\infty} \mathrm{d}t \, \frac{32G}{5c^5} \, \Izz^{2} \, \epsilon^2 \, \Omega^{6}(t) \,,
\end{equation}
must not exceed the remnant's initial rotational energy
\mbox{$\Erot = 0.5 \, \Izz \, \fref^2 \, \pi^2$}.

Given each pipeline's detection threshold,
we can rescale the amplitude of simulated signals
until \ULperc of them are recovered above threshold,
while randomising over nuisance parameters
(polarization angle and initial phase of the signal;
also source inclination $\iota$ and signal start time for some of the pipelines).
We can then either interpret this amplitude scaling
as a need to lower the distance of simulated sources,
i.e. estimating the sensitive distance $\dUL$ of the search.
Or we can fix the true distance to the source of \thebns
to obtain an energy upper limit $\EUL$:
For a remnant \ns with the given parameters,
we interpret the square of the amplitude scaling as the factor
by which the energy output needs to be greater than the expected amount
in order to produce signals that we can recover at 90\% confidence.
Both interpretations are shown in Fig.~\ref{fig:ULs_n5_dist_energy},
with coverage of the injection sets illustrated in Fig.~\ref{fig:UL_injs_coverage}.
Full results are listed in the Appendix in Tables \ref{tbl:ULs_n5_stamp}--\ref{tbl:ULs_n5_freqhough}.

The highest sensitivities are achieved at low $\fref$
and for rapid spindown (low $\tau$).
This is mostly due to the energy budget constraint
enforced on $\epsilon$:
In principle, higher $\fref$ yield higher initial amplitudes
and longer $\tau$ allow accumulation of \snr
over longer observation times
(\mbox{$\mathrm{\snr}\propto \tau^{1/2}$} for \mbox{$n = 5$} and once \mbox{$\Tobs > \tau$}).
However, due to the energy budget constraint
$\epsilon$ must be lower in this region of parameter space and hence actual detectability is reduced
(\mbox{$\mathrm{\snr}\propto\epsilon\propto\tau^{-1}\fref^{-6}$}).

The four pipelines perform differently across $\tau$ regimes:
The unmodeled \stamp and \hmm
are most sensitive at the shortest \mbox{$\tau=10^2$\,s},
but lose up to an order of magnitude in $\dUL$
when going to \mbox{$\tau=10^4$\,s}.
On the other hand,
the model-based semi-coherent \skyhough and \freqhough
have focused on longer $\tau$ of $4\times10^2$\,s to $3\times10^4$\,s,
with only up to a factor of 2 loss in $\dUL$ for the longest $\tau$ at fixed $\fref$.
See Figs. \ref{fig:ULs_n5_stamp}--\ref{fig:ULs_n5_freqhough} in the Appendix for sensitivity estimates
over each pipeline's full injection set.

As this parameter dependence is
shaped by the \mbox{$\Egw=\Erot$} constraint,
and also influenced by some practical tradeoffs in pipeline configuration,
in this paper we do not attempt to provide a general evaluation of pipeline performance
on fully equivalent injection sets,
nor for generic \gw signals.
Such a comparison would require a detailed mock data challenge similar to
\citet{Messenger:2015tga} and \citet{Walsh:2016hyc}.
Instead, Fig.~\ref{fig:ULs_n5_dist_energy}
shows results from each pipeline
for the parts of parameter space where it achieved its highest sensitivity.

In summary, in no part of the \mbox{$n=5$} parameter space
covered by the four search ranges and injection sets
do we reach \ULperc sensitive distances of \bestULdistance or further.
This corresponds to a lowest \ULperc detectable energy of
\mbox{$\Egw\lesssim\bestULenergy$}
at \mbox{$\fref=500$\,Hz} and \mbox{$\tau=100$\,s}.
At higher $\fref$, the sensitive distances for any $\tau$
are even lower
due to the energy constraint.
Note again that this covers power-law spindown signals
both starting right at coalescence time $\tc$
and signals starting with some time delay, with a delay time of 1--7\, hours for
\freqhough and any possible delay time until the end of O2 for \stamp.

At the shortest $\tau$, the parameter space covered here overlaps
with the magnetar injections in the shorter-duration search of \citet{Abbott:2017dke}\footnote{We
note here a mistake in \citet{Abbott:2017dke}:
In section 3.2.4, the equivalent energies for the best \stamp results
should have read
\mbox{$\Egw \approx 0.6 \Msun c^2$} for bar modes
and \mbox{$\Egw \approx 10 \Msun c^2$} for the magnetar model,
instead of the quoted 2 and 4 $\Msun c^2$.
The corresponding $\hrss$ values in the text of \citet{Abbott:2017dke}
and in its Tables 2 and 3,
as well as Figure~1, are correct as published.},
though results in that paper were quoted as recoverable at 50\% confidence,
and hence are more optimistic than the new results at \ULperc.
For example, at $\fref=1000$\,Hz and $\tau=100$\,s,
the \stamp analysis in the previous paper
found \mbox{$\Egw^{50\%}\approx24\,\Msun c^2$}
while the new \stamp and \hmm analyses presented here
obtain \mbox{$\Egw^{90\%}\approx100\,\Msun c^2$}
at these parameters.
For the pipelines in this paper, amplitudes for detectability at 50\% confidence are typically
lower by a factor of 2--4 than those at 90\% confidence.
While such lower thresholds would push the best $\olddUL$ limits up to a few Mpc,
this would not change the conclusion
that any \gw[s] from a long-lived remnant of \thebns at \distanceGWcompact would be undetectable.

\section{Conclusion}
\label{sec:conclusion}

We have searched for \gw emission from a putative remnant neutron star
of the \bns merger \thebns,
concentrating on signals lasting from hundreds of seconds upwards
and described by a power-law spin-down model.
Two of the four employed analysis methods
however were designed to be sensitive to any generic signal morphology
in the covered observation time.
In keeping with the available energy budget
and theoretical sensitivity estimates,
we have not found any significant signal candidates.
Studies with simulated signals confirm
that we would have only been sensitive to a signal
from \gw-dominated spin-down
(at the time and sky location of \thebns)
for distances of less than \bestULdistance,
or equivalently for unphysical amounts of emitted \gw energy.

The four analysis pipelines used in this work
have complementary strengths in parameter space coverage
and in their response to noise artifacts and gaps in the data.
While further development of these methods is expected,
improvements are also needed---and already in progress---on the instrumentation side.
Ongoing
instrumental enhancements of Advanced LIGO and Virgo
towards their design sensitivies~\citep{Aasi:2013wya},
and further upgrades like LIGO A+~\citep{T1800042} in the next decade,
will improve strain sensitivity across the detector band.
Improved high-frequency performance is of particular importance
for post-merger searches,
as the highest signal amplitudes are emitted
in the early, high-frequency part of the spin-down,
where the detectors are currently much less sensitive
than around a few hundred Hz.
Searches for long-duration post-merger signals
from supramassive or stable \ns[s]
could then enter into the astrophysically constraining regime.
However, from scaling the sensitivies
obtained in this analysis
\citep[or even those estimated for an optimal matched-filter analysis by][]{Sarin:2018vsi}
with the expected improvements of 2--4 in strain,
they will still be limited to the most nearby \bns events.

Third generation detectors, such as the
Einstein Telescope \citep{Hild:2010id,Sathyaprakash:2012jk}
and Cosmic Explorer \citep{Evans:2016mbw},
promise a strain sensitivity increase of \mbox{$\sim20$--$30$} over aLIGO at design sensitivity.
\gw[s] from a long-lived remnant of another \bns at the same distance as \thebns should then become observable.

\acknowledgments
\label{sec:ack}
%\bigskip\noindent\textit{Acknowledgments} ---
The authors gratefully acknowledge the support of the United States
National Science Foundation (NSF) for the construction and operation of the
LIGO Laboratory and Advanced LIGO as well as the Science and Technology Facilities Council (STFC) of the
United Kingdom, the Max-Planck-Society (MPS), and the State of
Niedersachsen/Germany for support of the construction of Advanced LIGO 
and construction and operation of the GEO600 detector. 
Additional support for Advanced LIGO was provided by the Australian Research Council.
The authors gratefully acknowledge the Italian Istituto Nazionale di Fisica Nucleare (INFN),  
the French Centre National de la Recherche Scientifique (CNRS) and
the Foundation for Fundamental Research on Matter supported by the Netherlands Organisation for Scientific Research, 
for the construction and operation of the Virgo detector
and the creation and support  of the EGO consortium. 
The authors also gratefully acknowledge research support from these agencies as well as by 
the Council of Scientific and Industrial Research of India, 
the Department of Science and Technology, India,
the Science \& Engineering Research Board (SERB), India,
the Ministry of Human Resource Development, India,
the Spanish  Agencia Estatal de Investigaci\'on,
the Vicepresid\`encia i Conselleria d'Innovaci\'o, Recerca i Turisme and the Conselleria d'Educaci\'o i Universitat del Govern de les Illes Balears,
the Conselleria d'Educaci\'o, Investigaci\'o, Cultura i Esport de la Generalitat Valenciana,
the National Science Centre of Poland,
the Swiss National Science Foundation (SNSF),
the Russian Foundation for Basic Research, 
the Russian Science Foundation,
the European Commission,
the European Regional Development Funds (ERDF),
the Royal Society, 
the Scottish Funding Council, 
the Scottish Universities Physics Alliance, 
the Hungarian Scientific Research Fund (OTKA),
the Lyon Institute of Origins (LIO),
the Paris \^{I}le-de-France Region, 
the National Research, Development and Innovation Office Hungary (NKFI), 
the National Research Foundation of Korea,
Industry Canada and the Province of Ontario through the Ministry of Economic Development and Innovation, 
the Natural Science and Engineering Research Council Canada,
the Canadian Institute for Advanced Research,
the Brazilian Ministry of Science, Technology, Innovations, and Communications,
the International Center for Theoretical Physics South American Institute for Fundamental Research (ICTP-SAIFR), 
the Research Grants Council of Hong Kong,
the National Natural Science Foundation of China (NSFC),
the Leverhulme Trust, 
the Research Corporation, 
the Ministry of Science and Technology (MOST), Taiwan
and
the Kavli Foundation.
The authors gratefully acknowledge the support of the NSF, STFC, MPS, INFN, CNRS and the
State of Niedersachsen/Germany for provision of computational resources.

This article has been assigned document number \dcc.

\clearpage

\appendix
\section{Additional details on search methods}
\label{sec:appendix-methods}

\subsection{STAMP}
\label{sec:appendix-methods-stamp}

\paragraph{Spectrogram pixel sizes}
The low frequency band from 30--2000\,Hz uses pixels of $100\,\mathrm{s} \times 1\,\mathrm{Hz}$,
while the high frequency band from 2000--4000\,Hz uses pixels of smaller durations of $50\,\mathrm{s} \times 1\,\mathrm{Hz}$.
Smaller pixels at higher frequency are necessary to account for the rotation of the Earth,
which causes the  \gw  phase difference between 
detectors to change with time. If the pixel durations are too large,
this results in a loss of \snr which increases with frequency.
The durations are thus chosen to limit the maximum possible \snr loss in a pixel
(at the highest frequencies)
from this effect to about $10\%$~\citep{Thrane:2015wla}.

\paragraph{Detection statistic}
Each spectrogram in the \stamp search~\citep{Thrane:2010ri,Thrane:2013bea,Thrane:2015wla}
is analyzed with many randomly chosen quadratic B{\'e}zier curves.
The SNR of each track $\rho_{\Gamma}$ is a weighted sum of the SNR of the pixels covered by the track.
The quantity $\rho_{\Gamma}$ also serves as the detection statistic and is calculated as:
\begin{equation}
\rho_{\Gamma} = \frac{1}{N^{3/4}} \sum_i \rho_i, 
\end{equation}
where $i$ runs over all the pixels in a track and N is the total number of pixels in it.
These are then ranked and the track with largest $\rho_{\Gamma}$
is picked as the trigger for a map.
This is done for both the main on-source search and for the background estimation over time-shifted data.

\paragraph{Background triggers and loudest events}
Fig.~\ref{fig:STAMP_fg_bg}
shows the distribution of false alarm probabilities $\pfa$
for the SNRs of triggers collected in background data,
for both high- and low-frequency spectrograms.
The loudest on-source event in each frequency range is also shown.

\begin{figure}
 \includegraphics{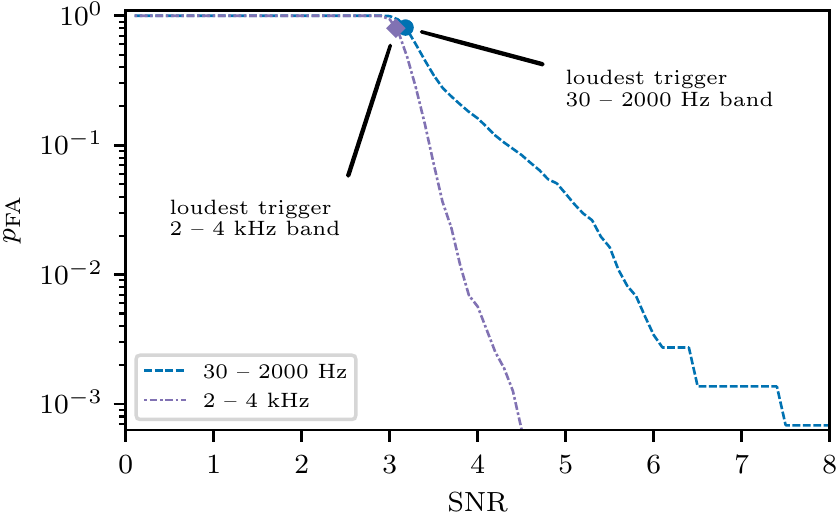}
 \caption{
  \label{fig:STAMP_fg_bg}
  \stamp background distributions,
  (in terms of false alarm probability $\pfa$
  as a function of detection statistic (SNR),
  for the low- and high-frequency bands,
  and the corresponding loudest foreground triggers
  (dot and diamond symbols).
 }
\end{figure}

\subsection{HMM tracking}
\label{sec:appendix-methods-hmm}

A general description of the \hmm method is given
by \citet{Suvorova:2016rdc} and \citet{Sun:2017zge}.
The following summary is intended to clarify the configuration
used for the search presented in this paper.

\paragraph{Probabilistic model}
A Markov chain is a stochastic process
transitioning between discrete states at discrete times
$\{t_0, \cdots, t_{N_T}\}$.
A HMM is an automaton based on a Markov chain,
composed of the hidden (unmeasurable) state variable
\mbox{$q(t) \in \{q_1, \cdots, q_{N_Q}\}$}
and the measurement variable
\mbox{$o(t)\in \{o_1, \cdots, o_{N_O}\}$}.
A HMM is memoryless,
i.e., the hidden state at time $t_{n+1}$
only depends on the state at time $t_n$,
with transition probability
\begin{equation}
 \label{eqn:prob_matrix}
 A_{q_j q_i} = P [q(t_{n+1})=q_j|q(t_n)=q_i].
\end{equation}
The hidden state $q_i$ is in observed state $o_j$ at time $t_n$
with emission probability
\begin{equation}
 \label{eqn:likelihood}
 L_{o_j q_i} = P [o(t_n)=o_j|q(t_n)=q_i].
\end{equation}
Given the prior defined by
\begin{equation}
 \Pi_{q_i} = P [q(t_0)=q_i],
\end{equation}
the probability that the hidden state path
\mbox{$Q=\{q(t_0), \cdots, q(t_{N_T})\}$}
gives rise to the observed sequence
\mbox{$O=\{o(t_0), \cdots, o(t_{N_T})\}$}
equals
\begin{equation}
 \label{eqn:prob}
 P(Q|O) =
 L_{o(t_{N_T})q(t_{N_T})} A_{q(t_{N_T})q(t_{N_T-1})} \cdots L_{o(t_1)q(t_1)}
 A_{q(t_1)q(t_0)} \Pi_{q(t_0)}.
\end{equation}
The most probable path
\begin{equation}
 Q^*(O)= \arg\max P(Q|O),
\end{equation}
maximizes $P(Q|O)$
and gives the best estimate of $q(t)$ over the total observation,
where $\arg \max (\cdots)$ returns the argument
that maximizes the function $(\cdots)$.
We use the classic Viterbi algorithm \citep{Viterbi:1967}
to efficiently solve the HMM and compute $Q^*(O)$. 

\paragraph{Search setup}
In this analysis, we track $q(t)=\fgw(t)$,
where $\fgw(t)$ is the \gw signal frequency at time $t$.
The discrete hidden states are mapped one-to-one
to the frequency bins in the output of a frequency-domain estimator $G(f)$
computed over an interval of length $\Tdrift$,
with bin size $\Delta f$.
We aim at searching for signals with $10^2\,{\rm s} \lesssim \tau \lesssim 10^4$\,s,
corresponding to $\fdotgw \lesssim 1$\,Hz\,s$^{-1}$.
We choose \mbox{$\Tdrift = 1$\,s}
(i.e., \mbox{$\Delta f = 1$\,Hz}) to satisfy
\begin{equation}
 \label{eqn:int_T_drift}
 \left|\int_t^{t+\Tdrift}dt' \fdotgw(t')\right| \leq \Delta f
\end{equation}
for \mbox{$0\leq t \leq T_{\rm obs}$}, where $T_{\rm obs}$ is the total observing time. 
The motion of the Earth with respect to the solar system barycenter (SSB)
can be neglected during the interval [$t,t+\Tdrift$].
Hence the emission probability
$L_{o(t)q_i} = P [o(t)|f_i \leq \fgw(t) \leq f_i+\Delta f]\propto \exp[G(f_i)]$
is calculated from the running-mean (window width 3\,Hz) normalized power
in \sft[s] with length
\mbox{$\Tsft = \Tdrift =1$\,s}
as the estimator $G(f)$.
We write
\begin{equation}
 G(f_i) = \sum_{X}\tilde{y}^X_i \tilde{y}^{X*}_i ,
\end{equation}
where $i$ indexes the frequency bins of the normalized SFT $\tilde{y}$, 
$X$ indexes the detector, and the repeated index $i$ on the right-hand side does not imply summation.
We assume that the auto-correlation time-scale of timing noise
is much longer than $\Tdrift$,
and hence adopt the transition probabilities
\begin{equation}
 \label{eqn:trans_matrix-pmr}
 A_{q_{i-1} q_i} = A_{q_i q_i} = \frac{1}{2},
\end{equation}
with all other entries being zero.
Since we have no independent knowledge of $f_{\rm gw}$,
we choose a uniform prior, viz.
\begin{equation}
 \Pi_{q_i} = N_Q^{-1}.
\end{equation}

We define a detection statistic $\mathcal{P}$, given by 
\begin{equation}
 \label{eqn:det_stat}
 \mathcal{P} = \frac{1}{N_T+1}\sum_{n=0}^{N_T} G[f_{i(t_n)}],
\end{equation}
where the integer $i(t_n)$ indexes the SFT frequency bin
corresponding to $q^*(t_n)$ on the optimal path $Q^*$ ($t_0\leq t_n \leq t_{N_T}$).

The strain amplitude $h_0$ in (\ref{eq:h0}) decreases significantly for $t \gg\tau$.
Hence the instant SNR decreases for $\Tobs \gtrsim \tau$. 
Monte-Carlo simulations show that choosing $\Tobs \sim \tau$
yields the best sensitivities for signals with $h_0$ (Eq.~\ref{eq:h0})
near the detection limit for the waveform in  Eq.~\ref{eq:fgw}.

The initial spin-down rate $|\fdotzero|$ of a signal with $\tau \lesssim 10^3$\,s can be too high
(i.e., $|\fdotzero|>1$\,Hz\,s$^{-1}$) for Eqn.~(\ref{eqn:int_T_drift}) to be satisfied with $\Tdrift=1$\,s. 
We start the search after waiting for a time $\twait$ post-merger, when $\fdotgw|$ decreases
such that Eqn.~(\ref{eqn:int_T_drift}) is satisfied.
Alternatively, we can choose shorter $\Tdrift$
(i.e., $\Tdrift \leq \fdotgw^{-1/2}$)
and take $\twait = 0$ for all waveforms.
However, the sensitivity degrades because the frequency resolution $\Delta f > 1$\,Hz
is relatively coarse for $\Tdrift<1$\,s.

In a search without prior knowledge of the signal model, we cover the following parameter space 
$500\,{\rm Hz} \leq \fzero \leq 2\,{\rm kHz}$ for
\mbox{$\tauminHMM\,{\rm s} \lesssim \tau \lesssim \taumaxHMM$\,s}
using seven discrete $\twait$ values in the range \mbox{$0 \leq \twait \leq 400$\,s}. 
Monte-Carlo simulations show that the impact on sensitivity from the mismatch 
in $\twait$ caused by the granularity is negligible.

\subsection{Adaptive Transient Hough}
\label{sec:appendix-methods-skyhough}

Here we summarize some key technical details concerning the search at hand,
while the complete derivation of the search method is in~\citet{Oliver:2019ksl}.

\paragraph{Coherence times}
The first step is selecting the coherent integration time $\Tcoh$,
i.e. the time-baseline of the \sft[s].
This cannot be arbitrarily large:
to avoid the signal shifting more than half a frequency bin,
$\Tcoh$ must satisfy $|\fdotgw|\Tcoh\leq 1/(2\Tcoh)$.
The time variation of $\fgw(t)$ is due to two effects: the spin-down of the source,
and the Doppler modulation due to the Earth's motion.
It is important to notice that in contrast to the continuous wave case
this method assumes that the Doppler modulation is a subdominant effect.
Thus $\Tcoh$ can be estimated as:
\begin{equation}
 \Tcoh \leq \frac{\sqrt{(n-1)\tau}}{\sqrt{2\fzero}}
\end{equation}

\paragraph{Hough transform}
Second, each of these \sft[s] is digitized by setting a threshold $\rho_\mathrm{th}$ on the normalized power,
that is directly related to the false alarm rate $\alpha$ and false dismissal rate $\beta$ of the search;
the optimal value is 1.6 as derived in~\citet{Krishnan:2004sv}.
This digitized spectrum is then weighted based on the noise floor of the detector
and the amplitude modulation of the source.
The derivation of the weights is given in~\citet{Oliver:2019ksl}
and in~\citet{T070124} for the CW all-sky case. 

\paragraph{Detection statistic and significance threshold}
Finally, each template---defined by
the set of signal parameters $(\fzero,n,\tau)$---is
incoherently integrated through the appropriate summation,
known as the number count,
over the weighted digitized spectrum following Eq.~\ref{eq:fgw}.
The critical ratio $\Psi$ is defined to evaluate the significance of a given template,
based on the results obtained for the weighted number count
and its estimates over Gaussian noise for the mean $\mu$ and the standard deviation $\sigma$:
\begin{equation}
 \Psi = \frac{n-\mu}{\sigma}
      = \frac{\sum\limits_{i=1}^{N_\mathrm{SFTs}}w_i y_i - \sum\limits_{i=1}^{N_\mathrm{SFTs}}w_i\alpha}{\sqrt{\sum\limits_{i=1}^{N_\mathrm{SFTs}}w_i^2\,\alpha\,(1-\alpha)}}
\end{equation}
Here, $y_i$ corresponds to the $ith$ digitalized bin in a given templated track, $N_\mathrm{SFTs}$ is the number of \sft[s] 
and the weights are \mbox{$w_i \propto (f_{\mathrm{gw},i})^{2m}(a_i^2+b_i^2)/S_{\mathrm{n},i}$},
where $a_i$ and $b_i$
are amplitude functions of the antenna pattern found in ~\citep{Jaranowski:1998qm}
at the $i^\mathrm{th}$ time step and $S_{\mathrm{n},i}$ is the power spectral density at that given bin.
As mentioned in~\citet{T070124} and \citet{Oliver:2019ksl},
any change in the normalization of the weights $w_i$ will not change the resulting sensitivity
and will also leave significances, or critical ratios $\Psi$, unchanged.
As shown in~\citet{Oliver:2019ksl}, the critical ratio for a multi-detector search can be written as
\begin{equation}
 \label{eq:ATrcriticalMultiSingle}
 \Psi_\mathrm{m} = \sum_{k=1}^{N_\mathrm{det}} \Psi_k r_{k}
\end{equation}
where $r_k$ is each detector's relative contribution ratio---proportional to
the number of \sft[s],
the noise power spectral density,
the antenna pattern and the signal amplitude pattern---
and $N_\mathrm{det}$ is the number of detectors.
For detection purposes, a threshold is placed on the two-detector $\Psi_\mathrm{m}$
corresponding to a $5\sigma$ false alarm probability for the entire template bank.
An additional single-detector threshold,
used for a consistency veto step,
is extrapolated from the $5\sigma$ threshold on $\Psi_\mathrm{m}$,
to make the veto safer under consideration of the actual differences in $r_k$ for this data set.

\paragraph{Candidate follow-up}
To verify that the \NcandsSkyHough outliers found in the initial search step were produced by noise in the detector,
and exclude the possiblility of having any actual astrophysical signals among them,
we performed an additional follow-up step.
For each template corresponding to one of the outliers,
an analoguous analysis was performed but with the template $\fgw(t)$ evolution starting instead 1\,hour after the merger.
From Eq.~\ref{eq:fgw}, templates starting at merger time have vanishing overlap with these +1\,hour delayed versions of themselves.
For all outliers, we find that the +1\,hour critical ratios are compatible
with the results found in the original search within $8\%$.
Given the $5\sigma$ false-alarm threshold imposed,
the critical ratio for these templates can be considered as
stationary non-Gaussian noise
with a false dismissal probability of less that $10^{-4}$.

\subsection{Generalized FrequencyHough}
\label{sec:appendix-methods-freqhough}

Full details of the adaptation of the FrequencyHough algorithm~\citep{Astone:2014esa}
to the case of rapid-spindown postmerger signals
are given in~\citet{Miller:2018genfreqhough}.
Here we summarize some technical details relevant to the
search presented in this paper.

\paragraph{Time offsets and search durations}
The search is run in three configurations using varying \mbox{$\Tfft=2,4,8$\,s},
covering different observing times:
$\toffset = 1$\,hour after merger with \mbox{$\Tfft=2$\,s} for signals lasting 700-7000\,s,
$\toffset \sim 1.5$--3\,hours after merger with \mbox{$\Tfft=4$\,s} for signals lasting 2000-16000\,s,
and $\toffset \sim 2$--7\,hours after merger with \mbox{$\Tfft=8$\,s} for signals lasting 8000-40000\,s.
The corresponding end times are set separately for each detector to guarantee that
the same effective amount of data is covered even in the presence of gaps;
the latest timestamps analysed for either detector are approximately 3, 8 and 22 hours after merger
in the three configurations.
See Fig.~\ref{fig:segs_FreqHough}.

\begin{figure}
 \includegraphics{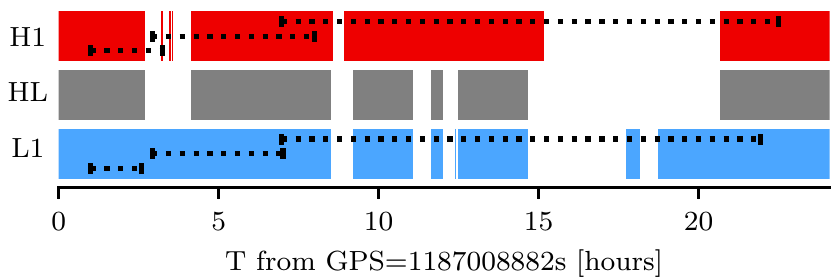}
 \caption{
  \label{fig:segs_FreqHough}
  \freqhough: Three configurations with different $\Tfft=2,4,8$ were used
  to search for a signal starting at least one hour to about 20 hours after the merger.
  The configurations were constructed to maximize sensitivity in different portions of the parameter space.
  We analyze the same amount of usable data in each detector, regardless of gaps in either detector's data,
  leading to later end times for the H1 detector in these three cases.
 }
\end{figure}

\paragraph{Coordinate transformation}
The parameters of the powerlaw spindown model
are transformed to new coordinates,
such that in the new space, the behavior of the signal is linear.
If we write $k'=(2\pi)^{n-1}k$,
we can rewrite Eq.~\ref{eq:fgw} as:
\begin{equation}
 \label{eq:freqhough_fgw}
 \fgw(t)=\frac{\fref}{\left[1 + k' (n-1)\fref^{n-1}(t-t_{ref})\right]^{\frac{1}{n-1}}}
\end{equation}
where:
\begin{equation}
 \label{eq:freqhough_k}
 k' =\frac{1}{\tau \fref^{n-1}(n-1)} \,.
\end{equation}
Then we can make the following change of coordinates:
\begin{equation}
 \label{eq:freqhough_transform}
 x=\frac{1}{\fgw^{n-1}}; \quad x_0=\frac{1}{\fref^{n-1}} \,.
\end{equation}
Equation \ref{eq:freqhough_fgw} becomes the equation of a line:
\begin{equation}
 \label{freqhough_xline}
 x=x_0+(n-1)k'(t-t_0)
\end{equation}
Now, we map peaks in the $(t-t_0,\fgw)$ plane of the detector
to lines in the ($x_0$,$k$) plane of the source,
where both variables in this space relate to the physical parameters of the source.

\paragraph{Grid setup}
Our method can be used to determine if a signal is present in the data and estimate its
$\fref$, $\fdotref$ and $n$.
We search across different braking indices with a fine, nonuniform grid
determined by ensuring that by stepping from $n$ to $n+dn$,
a signal is confined to one frequency bin.
For each braking index, we do a Hough transform and record the most significant candidates in the map.
The grid in $x_0$ is determined by taking the derivative of equation \ref{eq:freqhough_transform};
the grid in $k$ is created by considering a transformation
\mbox{$\fgw \rightarrow \, \fgw+ d\fgw=1/\Tfft$}
and \mbox{$k\rightarrow \, k+dk$}, then solving for $dk$ imposing that the spindown remains constant.

Both grids are nonuniform and depend on the frequency band and spindown range we use;
however, we overresolve the grid in $x_0$ for most frequency bands using the maximum frequency we are analyzing
because it is computationally faster and doesn't result in a sensitivity loss.

The transformation has the disadvantage that it creates nonuniform noise
(so peaks at higher frequencies contribute more to the Hough map).
We account for this by extending the frequency band we wish to analyze,
then only selecting candidates from the original band.

\paragraph{Coincidence step}
Candidates are considered in coincidence between detectors if the Euclidean distance
between their recovered parameters $x_0$ and $k$ is less than 3 bins.

\paragraph{Candidate follow-up}
Our candidate follow-up procedure is the following:
we correct for the phase evolution of the candidate recovered in each detector individually.
Ideally, if we correct for exactly the right parameters,
we expect a monochromatic signal in the time/frequency peakmap. If we are a bin or so off, there is a residual spindown or spinup, but the signal is linear.
Therefore, we can apply the original FrequencyHough to search for this signal.
After applying the original FrequencyHough
to the one surviving candidate from this search
and performing coincidences again,
we found no coincidence, indicating that the candidate was false.
We use the critical ratio as a way to veto candidates, defined as:

\begin{equation}
CR=\frac{y-\mu}{\sigma}
\end{equation}
where $y$ is the number count in a given bin in the Hough map,
$\mu$ is the average number count of the noise,
and $\sigma$ is the standard deviation of the number counts due to noise.
We determine if the critical ratio increases in the follow-up,
but find that it does not for our one remaining candidate.

\paragraph{Sensitivity estimation procedure}
We then computed the strain sensitivity for the different configuations of our search in the following way:
the loudest coincident candidate was selected in each 50 Hz band, for $n=5$.
Its $\fref$ and duration were used to inject signals with initial frequencies
ranging from $\fref$ to $\fref+50$ Hz,
with the highest possible spindown in our configuration $\fdotref=1/\Tfft^2$.
Based on our theoretical estimates for sensitivity,
the highest initial spindown corresponds to the most conservative result, i.e. the worst case sensitivity.
100 injections were done for each set of parameters,
i.e. each point plotted in Figs.~\ref{fig:ULs_n5_dist_energy}, \ref{fig:UL_injs_coverage} and \ref{fig:ULs_n5_freqhough}
and each row in \autoref{tbl:ULs_n5_freqhough}.
Recovery of an injection is defined in the same way as a dection in the real search:
within a coincidence distance of 3 bins.

\section{Details on pipeline sensitivity}
\label{sec:appendix-pipeline-sens}

In Figures \ref{fig:ULs_n5_stamp}--\ref{fig:ULs_n5_freqhough}
and Tables \ref{tbl:ULs_n5_stamp}--\ref{tbl:ULs_n5_freqhough}
we summarize the full sensitivity estimates
performed for each of the four pipelines
with their respective injection sets of simulated signals
following the power-law spin-down model (Eq.~\ref{eq:fgw})
with \mbox{$n=5$}.
Additional details on the injection sets
and sensitivity estimation procedure
for each pipeline
are given in the previous section or,
where necessary,
in the table captions.

\clearpage

\begin{figure*}
 \includegraphics{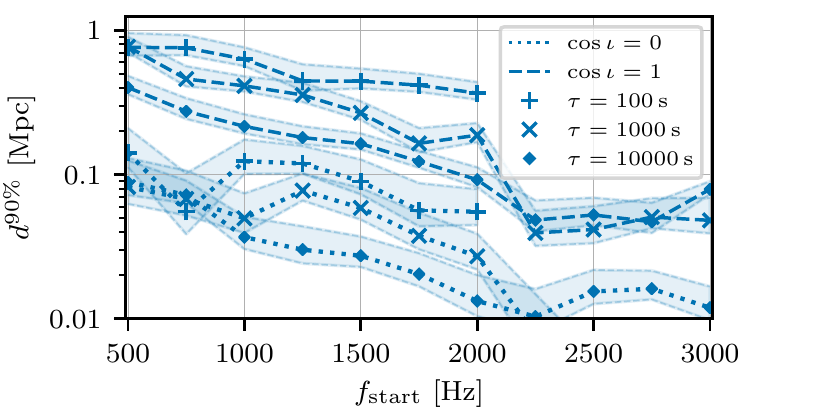}
 \includegraphics{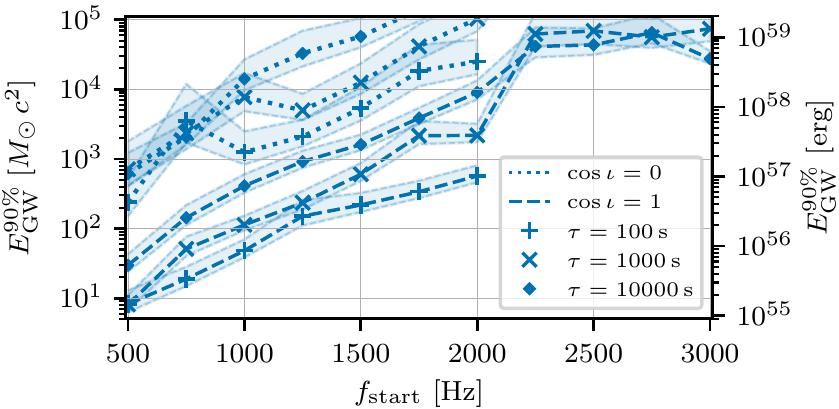}
 \caption{
  \label{fig:ULs_n5_stamp}
  \stamp \ULperc sensitivity estimates for $n=5$
  and variable \mbox{$\fref$}.
  For either $\cos\iota=[0,1]$,
  the connected lines (from top to bottom in $d$)
  are for injections with
  \mbox{$\tau \in [10^2,10^3,10^4]\,$s}.
  Shaded ranges correspond to $1\sigma$ binomial counting errors on the injection sets.
 }
\end{figure*}

\begin{figure*}
 \includegraphics{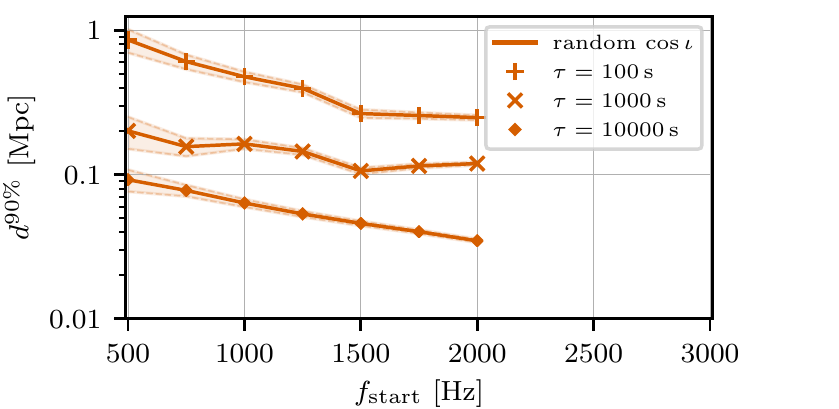}
 \includegraphics{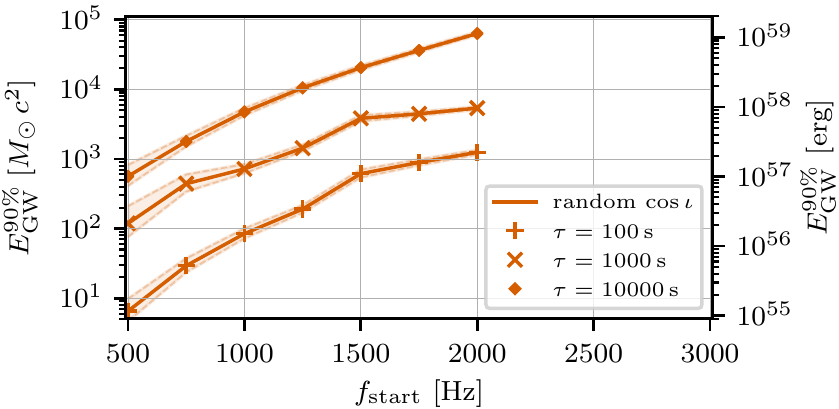}
 \caption{
  \label{fig:ULs_n5_hmm}
  \hmm \ULperc sensitivity estimates for $n=5$
  and \mbox{$\fref=\fzero=\fgw(\tc)$}.
  The connected lines (from top to bottom in $d$)
  are for injections with
  \mbox{$\tau \in [10^2,10^3,10^4]\,$s}.
  Shaded ranges illustrate the uncertainty due to interpolating
  discrete steps in injection amplitudes.
 }
\end{figure*}

\begin{figure*}
 \includegraphics{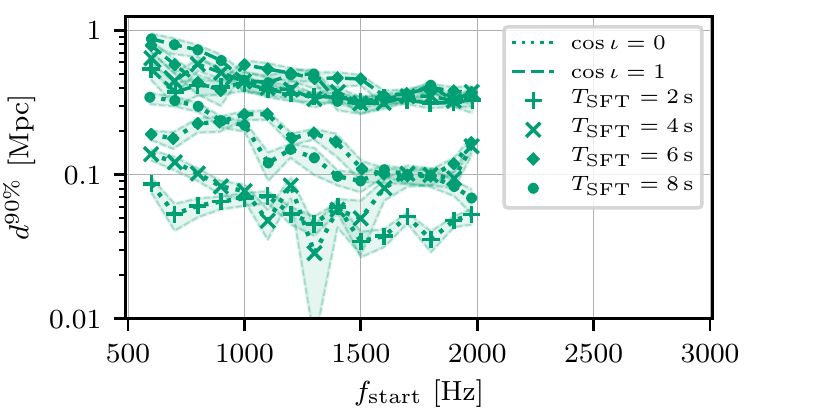}
 \includegraphics{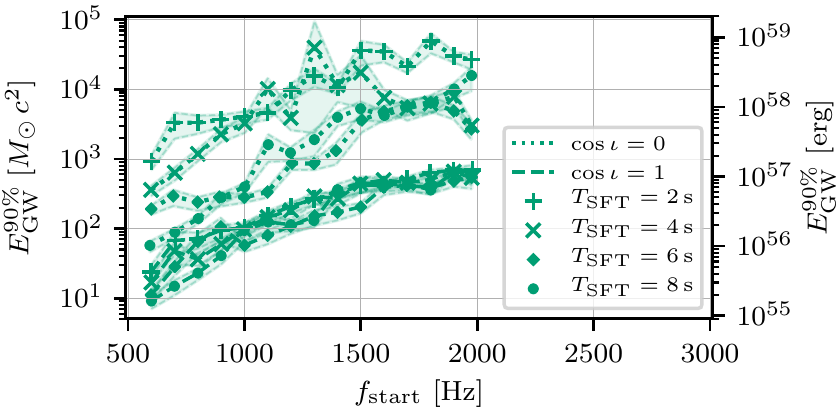}
 \caption{
  \label{fig:ULs_n5_skyhough}
  \skyhough \ULperc sensitivity estimates for $n=5$
  and \mbox{$\fref=\fzero=\fgw(\tc)$}.
  For either $\cos\iota=[0,1]$,
  the connected lines (from bottom to top in $d$)
  are for coherence times of $\Tsft=[2,4,6,8]$.
  Shaded ranges: 2$\sigma$ envelopes of logit fits over the different injection sets.
 }
\end{figure*}

\begin{figure*}
 \includegraphics{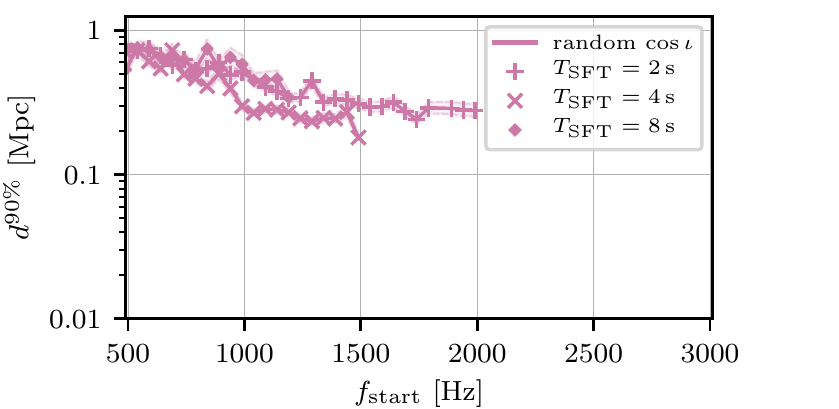}
 \includegraphics{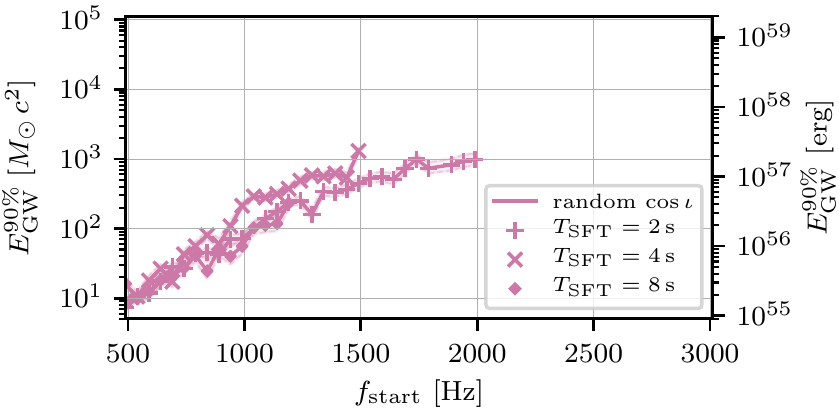}
 \caption{
  \label{fig:ULs_n5_freqhough}
  \freqhough \ULperc sensitivity estimates for $n=5$
  and \mbox{$\fref=\fgw(\tref=\tc+\toffset)$}.
  The connected lines
  are for coherence times of $\Tsft=[2,4,8]$\,s.
  Shaded ranges give the uncertainty due to interpolating
  discrete steps in injection amplitudes.
 }
\end{figure*}

\clearpage

\iftoggle{fulltables}{
 \startlongtable
}{}
\begin{deluxetable*}{crccll}
 \tablecaption{
  \label{tbl:ULs_n5_stamp}
  \stamp search sensitivities
  estimated from simulated signals (injections)
  following the power-law spindown model
  with braking index \mbox{$n=5$}.
  Each row corresponds to an injection set with fixed parameters;
  sensitivities are at \ULperc confidence.
 }
 \tablehead{
  \colhead{$\cos\iota$} &
  \colhead{$\fzero$\,[Hz]} &
  \colhead{$\tau$\, [s]} &
  \colhead{$\epsilon$} &
  \colhead{$\dUL$\,[Mpc]} &
  \colhead{$\EUL\,[\Msun c^2]$}
 }
 \startdata
  \iftoggle{fulltables}{
   0.0 &  500 & $1.00 \times 10^{2}$ & $7.33 \times 10^{-2}$ & $0.14_{-0.03}^{+0.07}$ & $2.39_{-0.88}^{+2.85} \times 10^{2}$\\
0.0 &  750 & $1.00 \times 10^{2}$ & $3.26 \times 10^{-2}$ & $0.055_{-0.017}^{+0.047}$ & $3.52_{-1.80}^{+8.43} \times 10^{3}$\\
0.0 & 1000 & $1.00 \times 10^{2}$ & $1.83 \times 10^{-2}$ & $0.12_{-0.02}^{+0.05}$ & $1.25_{-0.42}^{+1.25} \times 10^{3}$\\
0.0 & 1250 & $1.00 \times 10^{2}$ & $1.17 \times 10^{-2}$ & $0.12_{-0.02}^{+0.04}$ & $2.10_{-0.58}^{+1.57} \times 10^{3}$\\
0.0 & 1500 & $1.00 \times 10^{2}$ & $8.14 \times 10^{-3}$ & $0.090_{-0.017}^{+0.038}$ & $5.36_{-1.81}^{+5.45} \times 10^{3}$\\
0.0 & 1750 & $1.00 \times 10^{2}$ & $5.98 \times 10^{-3}$ & $0.057_{-0.013}^{+0.031}$ & $1.84_{-0.73}^{+2.53} \times 10^{4}$\\
0.0 & 2000 & $1.00 \times 10^{2}$ & $4.58 \times 10^{-3}$ & $0.055_{-0.011}^{+0.024}$ & $2.51_{-0.87}^{+2.65} \times 10^{4}$\\
0.0 &  500 & $1.00 \times 10^{3}$ & $2.32 \times 10^{-2}$ & $0.082_{-0.019}^{+0.048}$ & $7.11_{-2.97}^{+10.80} \times 10^{2}$\\
0.0 &  750 & $1.00 \times 10^{3}$ & $1.03 \times 10^{-2}$ & $0.068_{-0.016}^{+0.039}$ & $2.34_{-0.96}^{+3.43} \times 10^{3}$\\
0.0 & 1000 & $1.00 \times 10^{3}$ & $5.79 \times 10^{-3}$ & $0.050_{-0.010}^{+0.024}$ & $7.73_{-2.88}^{+9.40} \times 10^{3}$\\
0.0 & 1250 & $1.00 \times 10^{3}$ & $3.71 \times 10^{-3}$ & $0.077_{-0.011}^{+0.024}$ & $4.99_{-1.37}^{+3.64} \times 10^{3}$\\
0.0 & 1500 & $1.00 \times 10^{3}$ & $2.57 \times 10^{-3}$ & $0.059_{-0.010}^{+0.022}$ & $1.25_{-0.39}^{+1.11} \times 10^{4}$\\
0.0 & 1750 & $1.00 \times 10^{3}$ & $1.89 \times 10^{-3}$ & $0.038_{-0.007}^{+0.017}$ & $4.14_{-1.45}^{+4.51} \times 10^{4}$\\
0.0 & 2000 & $1.00 \times 10^{3}$ & $1.45 \times 10^{-3}$ & $0.027_{-0.005}^{+0.012}$ & $1.04_{-0.36}^{+1.09} \times 10^{5}$\\
0.0 & 2500 & $1.00 \times 10^{3}$ & $9.27 \times 10^{-4}$ & $0.0023_{-0.0009}^{+0.0034}$ & $2.34_{-1.55}^{+12.50} \times 10^{7}$\\
0.0 & 2750 & $1.00 \times 10^{3}$ & $7.66 \times 10^{-4}$ & $0.0036_{-0.0012}^{+0.0038}$ & $1.11_{-0.63}^{+3.52} \times 10^{7}$\\
0.0 & 3000 & $1.00 \times 10^{3}$ & $6.44 \times 10^{-4}$ & $0.0057_{-0.0015}^{+0.0037}$ & $5.32_{-2.37}^{+9.26} \times 10^{6}$\\
0.0 &  500 & $1.00 \times 10^{4}$ & $7.33 \times 10^{-3}$ & $0.088_{-0.016}^{+0.037}$ & $6.12_{-2.06}^{+6.15} \times 10^{2}$\\
0.0 &  750 & $1.00 \times 10^{4}$ & $3.26 \times 10^{-3}$ & $0.072_{-0.009}^{+0.019}$ & $2.06_{-0.48}^{+1.20} \times 10^{3}$\\
0.0 & 1000 & $1.00 \times 10^{4}$ & $1.83 \times 10^{-3}$ & $0.037_{-0.006}^{+0.014}$ & $1.41_{-0.45}^{+1.28} \times 10^{4}$\\
0.0 & 1250 & $1.00 \times 10^{4}$ & $1.17 \times 10^{-3}$ & $0.030_{-0.006}^{+0.014}$ & $3.29_{-1.17}^{+3.66} \times 10^{4}$\\
0.0 & 1500 & $1.00 \times 10^{4}$ & $8.14 \times 10^{-4}$ & $0.027_{-0.004}^{+0.010}$ & $5.75_{-1.73}^{+4.86} \times 10^{4}$\\
0.0 & 1750 & $1.00 \times 10^{4}$ & $5.98 \times 10^{-4}$ & $0.020_{-0.004}^{+0.008}$ & $1.41_{-0.46}^{+1.34} \times 10^{5}$\\
0.0 & 2000 & $1.00 \times 10^{4}$ & $4.58 \times 10^{-4}$ & $0.013_{-0.003}^{+0.007}$ & $4.34_{-1.67}^{+5.59} \times 10^{5}$\\
0.0 & 2250 & $1.00 \times 10^{4}$ & $3.62 \times 10^{-4}$ & $0.010_{-0.002}^{+0.006}$ & $9.02_{-3.63}^{+12.70} \times 10^{5}$\\
0.0 & 2500 & $1.00 \times 10^{4}$ & $2.93 \times 10^{-4}$ & $0.015_{-0.003}^{+0.006}$ & $4.99_{-1.64}^{+4.93} \times 10^{5}$\\
0.0 & 2750 & $1.00 \times 10^{4}$ & $2.42 \times 10^{-4}$ & $0.016_{-0.003}^{+0.005}$ & $5.55_{-1.60}^{+4.35} \times 10^{5}$\\
0.0 & 3000 & $1.00 \times 10^{4}$ & $2.04 \times 10^{-4}$ & $0.012_{-0.002}^{+0.005}$ & $1.21_{-0.40}^{+1.17} \times 10^{6}$\\
1.0 &  500 & $1.00 \times 10^{2}$ & $7.33 \times 10^{-2}$ & $0.76_{-0.10}^{+0.19}$ & $8.23_{-1.94}^{+4.72} \times 10^{0}$\\
1.0 &  750 & $1.00 \times 10^{2}$ & $3.26 \times 10^{-2}$ & $0.76_{-0.08}^{+0.17}$ & $1.88_{-0.40}^{+0.94} \times 10^{1}$\\
1.0 & 1000 & $1.00 \times 10^{2}$ & $1.83 \times 10^{-2}$ & $0.63_{-0.07}^{+0.13}$ & $4.80_{-0.96}^{+2.19} \times 10^{1}$\\
1.0 & 1250 & $1.00 \times 10^{2}$ & $1.17 \times 10^{-2}$ & $0.45_{-0.06}^{+0.14}$ & $1.51_{-0.40}^{+1.06} \times 10^{2}$\\
1.0 & 1500 & $1.00 \times 10^{2}$ & $8.14 \times 10^{-3}$ & $0.44_{-0.05}^{+0.10}$ & $2.18_{-0.45}^{+1.07} \times 10^{2}$\\
1.0 & 1750 & $1.00 \times 10^{2}$ & $5.98 \times 10^{-3}$ & $0.42_{-0.04}^{+0.08}$ & $3.38_{-0.65}^{+1.49} \times 10^{2}$\\
1.0 & 2000 & $1.00 \times 10^{2}$ & $4.58 \times 10^{-3}$ & $0.37_{-0.04}^{+0.07}$ & $5.69_{-1.07}^{+2.41} \times 10^{2}$\\
1.0 &  500 & $1.00 \times 10^{3}$ & $2.32 \times 10^{-2}$ & $0.77_{-0.07}^{+0.13}$ & $8.08_{-1.35}^{+2.92} \times 10^{0}$\\
1.0 &  750 & $1.00 \times 10^{3}$ & $1.03 \times 10^{-2}$ & $0.46_{-0.05}^{+0.10}$ & $5.09_{-1.08}^{+2.56} \times 10^{1}$\\
1.0 & 1000 & $1.00 \times 10^{3}$ & $5.79 \times 10^{-3}$ & $0.41_{-0.04}^{+0.07}$ & $1.13_{-0.19}^{+0.40} \times 10^{2}$\\
1.0 & 1250 & $1.00 \times 10^{3}$ & $3.71 \times 10^{-3}$ & $0.36_{-0.04}^{+0.08}$ & $2.35_{-0.48}^{+1.11} \times 10^{2}$\\
1.0 & 1500 & $1.00 \times 10^{3}$ & $2.57 \times 10^{-3}$ & $0.27_{-0.03}^{+0.05}$ & $6.03_{-1.19}^{+2.72} \times 10^{2}$\\
1.0 & 1750 & $1.00 \times 10^{3}$ & $1.89 \times 10^{-3}$ & $0.16_{-0.02}^{+0.05}$ & $2.16_{-0.54}^{+1.35} \times 10^{3}$\\
1.0 & 2000 & $1.00 \times 10^{3}$ & $1.45 \times 10^{-3}$ & $0.19_{-0.02}^{+0.04}$ & $2.18_{-0.44}^{+1.03} \times 10^{3}$\\
1.0 & 2250 & $1.00 \times 10^{3}$ & $1.14 \times 10^{-3}$ & $0.039_{-0.007}^{+0.017}$ & $6.24_{-2.12}^{+6.43} \times 10^{4}$\\
1.0 & 2500 & $1.00 \times 10^{3}$ & $9.27 \times 10^{-4}$ & $0.042_{-0.008}^{+0.019}$ & $6.92_{-2.45}^{+7.65} \times 10^{4}$\\
1.0 & 2750 & $1.00 \times 10^{3}$ & $7.66 \times 10^{-4}$ & $0.051_{-0.008}^{+0.018}$ & $5.62_{-1.69}^{+4.67} \times 10^{4}$\\
1.0 & 3000 & $1.00 \times 10^{3}$ & $6.44 \times 10^{-4}$ & $0.048_{-0.009}^{+0.021}$ & $7.43_{-2.52}^{+7.70} \times 10^{4}$\\
1.0 &  500 & $1.00 \times 10^{4}$ & $7.33 \times 10^{-3}$ & $0.40_{-0.04}^{+0.08}$ & $2.97_{-0.58}^{+1.33} \times 10^{1}$\\
1.0 &  750 & $1.00 \times 10^{4}$ & $3.26 \times 10^{-3}$ & $0.27_{-0.03}^{+0.06}$ & $1.43_{-0.31}^{+0.74} \times 10^{2}$\\
1.0 & 1000 & $1.00 \times 10^{4}$ & $1.83 \times 10^{-3}$ & $0.22_{-0.02}^{+0.05}$ & $4.11_{-0.84}^{+1.93} \times 10^{2}$\\
1.0 & 1250 & $1.00 \times 10^{4}$ & $1.17 \times 10^{-3}$ & $0.18_{-0.02}^{+0.04}$ & $9.18_{-1.74}^{+3.97} \times 10^{2}$\\
1.0 & 1500 & $1.00 \times 10^{4}$ & $8.14 \times 10^{-4}$ & $0.16_{-0.01}^{+0.03}$ & $1.61_{-0.27}^{+0.63} \times 10^{3}$\\
1.0 & 1750 & $1.00 \times 10^{4}$ & $5.98 \times 10^{-4}$ & $0.12_{-0.01}^{+0.02}$ & $3.86_{-0.70}^{+1.56} \times 10^{3}$\\
1.0 & 2000 & $1.00 \times 10^{4}$ & $4.58 \times 10^{-4}$ & $0.092_{-0.010}^{+0.020}$ & $9.04_{-1.90}^{+4.41} \times 10^{3}$\\
1.0 & 2250 & $1.00 \times 10^{4}$ & $3.62 \times 10^{-4}$ & $0.048_{-0.008}^{+0.018}$ & $4.14_{-1.26}^{+3.59} \times 10^{4}$\\
1.0 & 2500 & $1.00 \times 10^{4}$ & $2.93 \times 10^{-4}$ & $0.053_{-0.008}^{+0.017}$ & $4.34_{-1.21}^{+3.18} \times 10^{4}$\\
1.0 & 2750 & $1.00 \times 10^{4}$ & $2.42 \times 10^{-4}$ & $0.047_{-0.008}^{+0.017}$ & $6.54_{-1.96}^{+5.45} \times 10^{4}$\\
1.0 & 3000 & $1.00 \times 10^{4}$ & $2.04 \times 10^{-4}$ & $0.079_{-0.006}^{+0.011}$ & $2.75_{-0.41}^{+0.79} \times 10^{4}$\\

  }{
   0.0 &  500 & $1.00 \times 10^{2}$ & $7.33 \times 10^{-2}$ & $0.14_{-0.03}^{+0.07}$ & $2.39_{-0.88}^{+2.85} \times 10^{2}$\\
0.0 &  750 & $1.00 \times 10^{2}$ & $3.26 \times 10^{-2}$ & $0.055_{-0.017}^{+0.047}$ & $3.52_{-1.80}^{+8.43} \times 10^{3}$\\
0.0 & 1000 & $1.00 \times 10^{2}$ & $1.83 \times 10^{-2}$ & $0.12_{-0.02}^{+0.05}$ & $1.25_{-0.42}^{+1.25} \times 10^{3}$\\
0.0 & 1250 & $1.00 \times 10^{2}$ & $1.17 \times 10^{-2}$ & $0.12_{-0.02}^{+0.04}$ & $2.10_{-0.58}^{+1.57} \times 10^{3}$\\
0.0 & 1500 & $1.00 \times 10^{2}$ & $8.14 \times 10^{-3}$ & $0.090_{-0.017}^{+0.038}$ & $5.36_{-1.81}^{+5.45} \times 10^{3}$\\
\dots & \dots & \dots & \dots & \dots & \dots\\

  }
 \enddata
 \iftoggle{fulltables}{}{
  \tablecomments{Full table available as online machine-readable version.}
 }
\end{deluxetable*}

\iftoggle{fulltables}{
 \clearpage
 \startlongtable
}{}
\begin{deluxetable*}{rccll}
 \tablecaption{
  \label{tbl:ULs_n5_hmm}
  \hmm search sensitivities
  estimated from simulated signals (injections)
  following the power-law spindown model
  with braking index \mbox{$n=5$}.
  Each row corresponds to injections marginalized
  over random $\cos\iota$;
  sensitivities are at \ULperc confidence.
 }
 \tablehead{
  \colhead{$\fzero$\,[Hz]} &
  \colhead{$\tau$\, [s]} &
  \colhead{$\epsilon$} &
  \colhead{$\dUL$\,[Mpc]} &
  \colhead{$\EUL\,[\Msun c^2]$}
 }
 \startdata
  \iftoggle{fulltables}{
    500 & $1.00 \times 10^{2}$ & $7.33 \times 10^{-2}$ & $0.86_{-0.16}^{+0.16}$ & $6.51_{-1.87}^{+3.29} \times 10^{0}$\\
 750 & $1.00 \times 10^{2}$ & $3.26 \times 10^{-2}$ & $0.61_{-0.07}^{+0.07}$ & $2.92_{-0.58}^{+0.82} \times 10^{1}$\\
1000 & $1.00 \times 10^{2}$ & $1.83 \times 10^{-2}$ & $0.48_{-0.04}^{+0.04}$ & $8.43_{-1.25}^{+1.60} \times 10^{1}$\\
1250 & $1.00 \times 10^{2}$ & $1.17 \times 10^{-2}$ & $0.40_{-0.03}^{+0.03}$ & $1.90_{-0.22}^{+0.27} \times 10^{2}$\\
1500 & $1.00 \times 10^{2}$ & $8.14 \times 10^{-3}$ & $0.26_{-0.02}^{+0.02}$ & $6.15_{-0.75}^{+0.91} \times 10^{2}$\\
1750 & $1.00 \times 10^{2}$ & $5.98 \times 10^{-3}$ & $0.26_{-0.01}^{+0.01}$ & $8.90_{-0.83}^{+0.97} \times 10^{2}$\\
2000 & $1.00 \times 10^{2}$ & $4.58 \times 10^{-3}$ & $0.25_{-0.01}^{+0.01}$ & $1.24_{-0.09}^{+0.11} \times 10^{3}$\\
 500 & $1.00 \times 10^{3}$ & $2.32 \times 10^{-2}$ & $0.20_{-0.05}^{+0.05}$ & $1.19_{-0.43}^{+0.92} \times 10^{2}$\\
 750 & $1.00 \times 10^{3}$ & $1.03 \times 10^{-2}$ & $0.16_{-0.02}^{+0.02}$ & $4.41_{-1.03}^{+1.59} \times 10^{2}$\\
1000 & $1.00 \times 10^{3}$ & $5.79 \times 10^{-3}$ & $0.16_{-0.01}^{+0.01}$ & $7.19_{-0.99}^{+1.25} \times 10^{2}$\\
1250 & $1.00 \times 10^{3}$ & $3.71 \times 10^{-3}$ & $0.14_{-0.01}^{+0.01}$ & $1.43_{-0.15}^{+0.17} \times 10^{3}$\\
1500 & $1.00 \times 10^{3}$ & $2.57 \times 10^{-3}$ & $0.11_{-0.01}^{+0.01}$ & $3.83_{-0.37}^{+0.44} \times 10^{3}$\\
1750 & $1.00 \times 10^{3}$ & $1.89 \times 10^{-3}$ & $0.11_{-0.004}^{+0.004}$ & $4.45_{-0.30}^{+0.34} \times 10^{3}$\\
2000 & $1.00 \times 10^{3}$ & $1.45 \times 10^{-3}$ & $0.12_{-0.003}^{+0.003}$ & $5.38_{-0.27}^{+0.30} \times 10^{3}$\\
 500 & $1.00 \times 10^{4}$ & $7.33 \times 10^{-3}$ & $0.092_{-0.016}^{+0.016}$ & $5.64_{-1.54}^{+2.60} \times 10^{2}$\\
 750 & $1.00 \times 10^{4}$ & $3.26 \times 10^{-3}$ & $0.078_{-0.007}^{+0.007}$ & $1.79_{-0.29}^{+0.38} \times 10^{3}$\\
1000 & $1.00 \times 10^{4}$ & $1.83 \times 10^{-3}$ & $0.064_{-0.004}^{+0.004}$ & $4.74_{-0.54}^{+0.65} \times 10^{3}$\\
1250 & $1.00 \times 10^{4}$ & $1.17 \times 10^{-3}$ & $0.053_{-0.003}^{+0.003}$ & $1.05_{-0.09}^{+0.11} \times 10^{4}$\\
1500 & $1.00 \times 10^{4}$ & $8.14 \times 10^{-4}$ & $0.046_{-0.002}^{+0.002}$ & $2.05_{-0.15}^{+0.17} \times 10^{4}$\\
1750 & $1.00 \times 10^{4}$ & $5.98 \times 10^{-4}$ & $0.040_{-0.001}^{+0.001}$ & $3.63_{-0.22}^{+0.25} \times 10^{4}$\\
2000 & $1.00 \times 10^{4}$ & $4.58 \times 10^{-4}$ & $0.035_{-0.001}^{+0.001}$ & $6.35_{-0.35}^{+0.38} \times 10^{4}$\\

  }{
    500 & $1.00 \times 10^{2}$ & $7.33 \times 10^{-2}$ & $0.86_{-0.16}^{+0.16}$ & $6.51_{-1.87}^{+3.29} \times 10^{0}$\\
 750 & $1.00 \times 10^{2}$ & $3.26 \times 10^{-2}$ & $0.61_{-0.07}^{+0.07}$ & $2.92_{-0.58}^{+0.82} \times 10^{1}$\\
1000 & $1.00 \times 10^{2}$ & $1.83 \times 10^{-2}$ & $0.48_{-0.04}^{+0.04}$ & $8.43_{-1.25}^{+1.60} \times 10^{1}$\\
1250 & $1.00 \times 10^{2}$ & $1.17 \times 10^{-2}$ & $0.40_{-0.03}^{+0.03}$ & $1.90_{-0.22}^{+0.27} \times 10^{2}$\\
1500 & $1.00 \times 10^{2}$ & $8.14 \times 10^{-3}$ & $0.26_{-0.02}^{+0.02}$ & $6.15_{-0.75}^{+0.91} \times 10^{2}$\\
\dots & \dots & \dots & \dots & \dots\\

  }
 \enddata
 \iftoggle{fulltables}{}{
  \tablecomments{Full table available as online machine-readable version.}
 }
\end{deluxetable*}

\iftoggle{fulltables}{
 \clearpage
 \startlongtable
}{}
\begin{deluxetable*}{ccrccll}
 \tablecaption{
  \label{tbl:ULs_n5_skyhough}
  \skyhough search sensitivities
  estimated from simulated signals (injections)
  following the power-law spindown model
  with braking index \mbox{$n=5$}.
  Each row corresponds to injections marginalized
  over a small band in $\fzero$;
  sensitivities are at \ULperc confidence.
 }
 \tablehead{
  \colhead{$\Tsft$\,[s]} &
  \colhead{$\cos\iota$} &
  \colhead{$\fzero$\,[Hz]} &
  \colhead{$\tau$\, [s]} &
  \colhead{$\epsilon$} &
  \colhead{$\dUL$\,[Mpc]} &
  \colhead{$\EUL\,[\Msun c^2]$}
 }
 \startdata
  \iftoggle{fulltables}{
   2 & 0.0 &  601 & $6.01 \times 10^{2}$ & $2.07 \times 10^{-2}$ & $0.086_{-0.012}^{+0.009}$ & $9.30_{-2.32}^{+1.96} \times 10^{2}$\\
2 & 0.0 &  700 & $7.00 \times 10^{2}$ & $1.41 \times 10^{-2}$ & $0.053_{-0.012}^{+0.010}$ & $3.32_{-1.37}^{+1.29} \times 10^{3}$\\
2 & 0.0 &  800 & $8.00 \times 10^{2}$ & $1.01 \times 10^{-2}$ & $0.061_{-0.010}^{+0.008}$ & $3.32_{-1.03}^{+0.89} \times 10^{3}$\\
2 & 0.0 &  898 & $8.98 \times 10^{2}$ & $7.57 \times 10^{-3}$ & $0.065_{-0.008}^{+0.006}$ & $3.66_{-0.81}^{+0.65} \times 10^{3}$\\
2 & 0.0 & 1000 & $1.00 \times 10^{3}$ & $5.80 \times 10^{-3}$ & $0.069_{-0.008}^{+0.006}$ & $4.07_{-0.89}^{+0.73} \times 10^{3}$\\
2 & 0.0 & 1100 & $1.10 \times 10^{3}$ & $4.56 \times 10^{-3}$ & $0.071_{-0.008}^{+0.006}$ & $4.60_{-0.94}^{+0.78} \times 10^{3}$\\
2 & 0.0 & 1200 & $1.20 \times 10^{3}$ & $3.67 \times 10^{-3}$ & $0.053_{-0.008}^{+0.006}$ & $9.75_{-2.80}^{+2.36} \times 10^{3}$\\
2 & 0.0 & 1300 & $1.30 \times 10^{3}$ & $3.01 \times 10^{-3}$ & $0.046_{-0.008}^{+0.006}$ & $1.56_{-0.48}^{+0.42} \times 10^{4}$\\
2 & 0.0 & 1401 & $1.40 \times 10^{3}$ & $2.49 \times 10^{-3}$ & $0.060_{-0.006}^{+0.004}$ & $1.05_{-0.19}^{+0.15} \times 10^{4}$\\
2 & 0.0 & 1500 & $1.50 \times 10^{3}$ & $2.10 \times 10^{-3}$ & $0.034_{-0.008}^{+0.006}$ & $3.63_{-1.44}^{+1.29} \times 10^{4}$\\
2 & 0.0 & 1600 & $1.60 \times 10^{3}$ & $1.79 \times 10^{-3}$ & $0.037_{-0.006}^{+0.005}$ & $3.49_{-1.05}^{+0.89} \times 10^{4}$\\
2 & 0.0 & 1700 & $1.70 \times 10^{3}$ & $1.54 \times 10^{-3}$ & $0.051_{-0.005}^{+0.004}$ & $2.13_{-0.41}^{+0.31} \times 10^{4}$\\
2 & 0.0 & 1801 & $1.80 \times 10^{3}$ & $1.33 \times 10^{-3}$ & $0.036_{-0.007}^{+0.005}$ & $4.93_{-1.65}^{+1.47} \times 10^{4}$\\
2 & 0.0 & 1900 & $1.90 \times 10^{3}$ & $1.16 \times 10^{-3}$ & $0.048_{-0.005}^{+0.003}$ & $3.01_{-0.57}^{+0.45} \times 10^{4}$\\
2 & 0.0 & 1974 & $1.97 \times 10^{3}$ & $1.06 \times 10^{-3}$ & $0.053_{-0.008}^{+0.005}$ & $2.67_{-0.75}^{+0.49} \times 10^{4}$\\
4 & 0.0 &  599 & $2.40 \times 10^{3}$ & $1.04 \times 10^{-2}$ & $0.14_{-0.01}^{+0.01}$ & $3.59_{-0.65}^{+0.50} \times 10^{2}$\\
4 & 0.0 &  701 & $2.80 \times 10^{3}$ & $7.04 \times 10^{-3}$ & $0.12_{-0.01}^{+0.01}$ & $6.35_{-1.46}^{+1.20} \times 10^{2}$\\
4 & 0.0 &  800 & $3.20 \times 10^{3}$ & $5.06 \times 10^{-3}$ & $0.10_{-0.01}^{+0.01}$ & $1.18_{-0.26}^{+0.21} \times 10^{3}$\\
4 & 0.0 &  899 & $3.60 \times 10^{3}$ & $3.78 \times 10^{-3}$ & $0.083_{-0.010}^{+0.008}$ & $2.26_{-0.53}^{+0.43} \times 10^{3}$\\
4 & 0.0 & 1001 & $4.00 \times 10^{3}$ & $2.89 \times 10^{-3}$ & $0.077_{-0.011}^{+0.009}$ & $3.25_{-0.89}^{+0.76} \times 10^{3}$\\
4 & 0.0 & 1100 & $4.40 \times 10^{3}$ & $2.28 \times 10^{-3}$ & $0.048_{-0.012}^{+0.009}$ & $1.01_{-0.46}^{+0.43} \times 10^{4}$\\
4 & 0.0 & 1199 & $4.80 \times 10^{3}$ & $1.84 \times 10^{-3}$ & $0.084_{-0.016}^{+0.011}$ & $3.90_{-1.32}^{+1.14} \times 10^{3}$\\
4 & 0.0 & 1301 & $5.20 \times 10^{3}$ & $1.50 \times 10^{-3}$ & $0.029_{-0.022}^{+0.016}$ & $3.97_{-3.74}^{+5.55} \times 10^{4}$\\
4 & 0.0 & 1400 & $5.60 \times 10^{3}$ & $1.25 \times 10^{-3}$ & $0.057_{-0.014}^{+0.011}$ & $1.15_{-0.50}^{+0.47} \times 10^{4}$\\
4 & 0.0 & 1500 & $6.00 \times 10^{3}$ & $1.05 \times 10^{-3}$ & $0.050_{-0.022}^{+0.016}$ & $1.74_{-1.20}^{+1.32} \times 10^{4}$\\
4 & 0.0 & 1601 & $6.40 \times 10^{3}$ & $8.93 \times 10^{-4}$ & $0.081_{-0.015}^{+0.011}$ & $7.56_{-2.50}^{+2.24} \times 10^{3}$\\
4 & 0.0 & 1700 & $6.80 \times 10^{3}$ & $7.69 \times 10^{-4}$ & $0.10_{-0.02}^{+0.01}$ & $5.38_{-1.83}^{+1.62} \times 10^{3}$\\
4 & 0.0 & 1799 & $7.20 \times 10^{3}$ & $6.67 \times 10^{-4}$ & $0.10_{-0.02}^{+0.01}$ & $6.26_{-1.76}^{+1.49} \times 10^{3}$\\
4 & 0.0 & 1901 & $7.60 \times 10^{3}$ & $5.81 \times 10^{-4}$ & $0.094_{-0.014}^{+0.010}$ & $7.82_{-2.11}^{+1.76} \times 10^{3}$\\
4 & 0.0 & 1976 & $7.90 \times 10^{3}$ & $5.28 \times 10^{-4}$ & $0.16_{-0.02}^{+0.01}$ & $3.02_{-0.78}^{+0.56} \times 10^{3}$\\
6 & 0.0 &  600 & $5.40 \times 10^{3}$ & $6.93 \times 10^{-3}$ & $0.19_{-0.02}^{+0.01}$ & $1.91_{-0.32}^{+0.23} \times 10^{2}$\\
6 & 0.0 &  694 & $6.25 \times 10^{3}$ & $4.81 \times 10^{-3}$ & $0.18_{-0.03}^{+0.02}$ & $2.92_{-0.82}^{+0.67} \times 10^{2}$\\
6 & 0.0 &  799 & $7.19 \times 10^{3}$ & $3.38 \times 10^{-3}$ & $0.23_{-0.03}^{+0.02}$ & $2.40_{-0.60}^{+0.43} \times 10^{2}$\\
6 & 0.0 &  892 & $8.03 \times 10^{3}$ & $2.57 \times 10^{-3}$ & $0.23_{-0.03}^{+0.02}$ & $2.84_{-0.78}^{+0.53} \times 10^{2}$\\
6 & 0.0 & 1000 & $9.00 \times 10^{3}$ & $1.93 \times 10^{-3}$ & $0.26_{-0.02}^{+0.01}$ & $2.80_{-0.45}^{+0.29} \times 10^{2}$\\
6 & 0.0 & 1099 & $9.89 \times 10^{3}$ & $1.53 \times 10^{-3}$ & $0.26_{-0.02}^{+0.02}$ & $3.35_{-0.57}^{+0.41} \times 10^{2}$\\
6 & 0.0 & 1204 & $1.08 \times 10^{4}$ & $1.21 \times 10^{-3}$ & $0.18_{-0.02}^{+0.01}$ & $8.69_{-1.74}^{+1.39} \times 10^{2}$\\
6 & 0.0 & 1298 & $1.17 \times 10^{4}$ & $1.01 \times 10^{-3}$ & $0.19_{-0.02}^{+0.01}$ & $8.57_{-1.63}^{+1.25} \times 10^{2}$\\
6 & 0.0 & 1393 & $1.25 \times 10^{4}$ & $8.43 \times 10^{-4}$ & $0.17_{-0.03}^{+0.02}$ & $1.31_{-0.48}^{+0.39} \times 10^{3}$\\
6 & 0.0 & 1505 & $1.35 \times 10^{4}$ & $6.95 \times 10^{-4}$ & $0.11_{-0.02}^{+0.01}$ & $3.63_{-1.22}^{+1.05} \times 10^{3}$\\
6 & 0.0 & 1600 & $1.44 \times 10^{4}$ & $5.97 \times 10^{-4}$ & $0.10_{-0.01}^{+0.01}$ & $4.82_{-1.22}^{+1.02} \times 10^{3}$\\
6 & 0.0 & 1700 & $1.53 \times 10^{4}$ & $5.12 \times 10^{-4}$ & $0.098_{-0.012}^{+0.009}$ & $5.78_{-1.32}^{+1.09} \times 10^{3}$\\
6 & 0.0 & 1800 & $1.62 \times 10^{4}$ & $4.44 \times 10^{-4}$ & $0.098_{-0.014}^{+0.010}$ & $6.49_{-1.72}^{+1.42} \times 10^{3}$\\
6 & 0.0 & 1900 & $1.71 \times 10^{4}$ & $3.88 \times 10^{-4}$ & $0.12_{-0.02}^{+0.01}$ & $4.88_{-1.18}^{+1.00} \times 10^{3}$\\
6 & 0.0 & 1974 & $1.78 \times 10^{4}$ & $3.52 \times 10^{-4}$ & $0.17_{-0.03}^{+0.02}$ & $2.70_{-0.79}^{+0.51} \times 10^{3}$\\
8 & 0.0 &  594 & $9.50 \times 10^{3}$ & $5.33 \times 10^{-3}$ & $0.34_{-0.04}^{+0.02}$ & $5.72_{-1.12}^{+0.83} \times 10^{1}$\\
8 & 0.0 &  701 & $1.12 \times 10^{4}$ & $3.52 \times 10^{-3}$ & $0.33_{-0.03}^{+0.02}$ & $8.87_{-1.54}^{+1.12} \times 10^{1}$\\
8 & 0.0 &  802 & $1.28 \times 10^{4}$ & $2.52 \times 10^{-3}$ & $0.30_{-0.03}^{+0.02}$ & $1.39_{-0.23}^{+0.15} \times 10^{2}$\\
8 & 0.0 &  901 & $1.44 \times 10^{4}$ & $1.88 \times 10^{-3}$ & $0.24_{-0.02}^{+0.02}$ & $2.80_{-0.55}^{+0.40} \times 10^{2}$\\
8 & 0.0 & 1001 & $1.60 \times 10^{4}$ & $1.44 \times 10^{-3}$ & $0.22_{-0.02}^{+0.01}$ & $3.98_{-0.73}^{+0.53} \times 10^{2}$\\
8 & 0.0 & 1102 & $1.76 \times 10^{4}$ & $1.13 \times 10^{-3}$ & $0.12_{-0.03}^{+0.02}$ & $1.60_{-0.70}^{+0.65} \times 10^{3}$\\
8 & 0.0 & 1198 & $1.92 \times 10^{4}$ & $9.22 \times 10^{-4}$ & $0.15_{-0.02}^{+0.01}$ & $1.23_{-0.29}^{+0.23} \times 10^{3}$\\
8 & 0.0 & 1300 & $2.08 \times 10^{4}$ & $7.51 \times 10^{-4}$ & $0.13_{-0.03}^{+0.02}$ & $1.90_{-0.80}^{+0.71} \times 10^{3}$\\
8 & 0.0 & 1400 & $2.24 \times 10^{4}$ & $6.24 \times 10^{-4}$ & $0.097_{-0.013}^{+0.010}$ & $3.97_{-1.00}^{+0.85} \times 10^{3}$\\
8 & 0.0 & 1499 & $2.40 \times 10^{4}$ & $5.26 \times 10^{-4}$ & $0.090_{-0.015}^{+0.011}$ & $5.27_{-1.60}^{+1.39} \times 10^{3}$\\
8 & 0.0 & 1601 & $2.56 \times 10^{4}$ & $4.46 \times 10^{-4}$ & $0.11_{-0.01}^{+0.01}$ & $4.21_{-0.84}^{+0.66} \times 10^{3}$\\
8 & 0.0 & 1700 & $2.72 \times 10^{4}$ & $3.84 \times 10^{-4}$ & $0.10_{-0.01}^{+0.01}$ & $5.57_{-1.37}^{+1.14} \times 10^{3}$\\
8 & 0.0 & 1799 & $2.88 \times 10^{4}$ & $3.34 \times 10^{-4}$ & $0.098_{-0.016}^{+0.012}$ & $6.45_{-1.93}^{+1.68} \times 10^{3}$\\
8 & 0.0 & 1901 & $3.04 \times 10^{4}$ & $2.91 \times 10^{-4}$ & $0.083_{-0.012}^{+0.009}$ & $1.01_{-0.27}^{+0.22} \times 10^{4}$\\
8 & 0.0 & 1976 & $3.16 \times 10^{4}$ & $2.64 \times 10^{-4}$ & $0.069_{-0.016}^{+0.010}$ & $1.58_{-0.63}^{+0.50} \times 10^{4}$\\
2 & 1.0 &  599 & $5.99 \times 10^{2}$ & $2.09 \times 10^{-2}$ & $0.54_{-0.08}^{+0.06}$ & $2.38_{-0.68}^{+0.58} \times 10^{1}$\\
2 & 1.0 &  703 & $7.03 \times 10^{2}$ & $1.40 \times 10^{-2}$ & $0.37_{-0.08}^{+0.06}$ & $6.76_{-2.52}^{+2.21} \times 10^{1}$\\
2 & 1.0 &  799 & $7.99 \times 10^{2}$ & $1.02 \times 10^{-2}$ & $0.41_{-0.06}^{+0.05}$ & $7.11_{-1.98}^{+1.65} \times 10^{1}$\\
2 & 1.0 &  900 & $9.00 \times 10^{2}$ & $7.55 \times 10^{-3}$ & $0.41_{-0.05}^{+0.03}$ & $9.42_{-2.10}^{+1.69} \times 10^{1}$\\
2 & 1.0 & 1001 & $1.00 \times 10^{3}$ & $5.78 \times 10^{-3}$ & $0.43_{-0.05}^{+0.03}$ & $1.04_{-0.22}^{+0.17} \times 10^{2}$\\
2 & 1.0 & 1101 & $1.10 \times 10^{3}$ & $4.55 \times 10^{-3}$ & $0.39_{-0.05}^{+0.03}$ & $1.53_{-0.34}^{+0.28} \times 10^{2}$\\
2 & 1.0 & 1199 & $1.20 \times 10^{3}$ & $3.68 \times 10^{-3}$ & $0.36_{-0.04}^{+0.03}$ & $2.09_{-0.44}^{+0.33} \times 10^{2}$\\
2 & 1.0 & 1299 & $1.30 \times 10^{3}$ & $3.01 \times 10^{-3}$ & $0.35_{-0.03}^{+0.02}$ & $2.66_{-0.46}^{+0.33} \times 10^{2}$\\
2 & 1.0 & 1402 & $1.40 \times 10^{3}$ & $2.49 \times 10^{-3}$ & $0.34_{-0.03}^{+0.02}$ & $3.20_{-0.56}^{+0.43} \times 10^{2}$\\
2 & 1.0 & 1500 & $1.50 \times 10^{3}$ & $2.10 \times 10^{-3}$ & $0.32_{-0.03}^{+0.02}$ & $4.15_{-0.85}^{+0.65} \times 10^{2}$\\
2 & 1.0 & 1599 & $1.60 \times 10^{3}$ & $1.79 \times 10^{-3}$ & $0.34_{-0.03}^{+0.02}$ & $4.24_{-0.66}^{+0.47} \times 10^{2}$\\
2 & 1.0 & 1699 & $1.70 \times 10^{3}$ & $1.54 \times 10^{-3}$ & $0.33_{-0.02}^{+0.01}$ & $5.14_{-0.67}^{+0.44} \times 10^{2}$\\
2 & 1.0 & 1798 & $1.80 \times 10^{3}$ & $1.34 \times 10^{-3}$ & $0.31_{-0.02}^{+0.02}$ & $6.35_{-0.94}^{+0.64} \times 10^{2}$\\
2 & 1.0 & 1897 & $1.90 \times 10^{3}$ & $1.17 \times 10^{-3}$ & $0.32_{-0.02}^{+0.01}$ & $6.90_{-0.87}^{+0.58} \times 10^{2}$\\
2 & 1.0 & 1979 & $1.98 \times 10^{3}$ & $1.05 \times 10^{-3}$ & $0.33_{-0.06}^{+0.03}$ & $6.92_{-2.37}^{+1.22} \times 10^{2}$\\
4 & 1.0 &  600 & $2.40 \times 10^{3}$ & $1.04 \times 10^{-2}$ & $0.64_{-0.09}^{+0.06}$ & $1.68_{-0.43}^{+0.35} \times 10^{1}$\\
4 & 1.0 &  700 & $2.80 \times 10^{3}$ & $7.08 \times 10^{-3}$ & $0.44_{-0.09}^{+0.07}$ & $4.75_{-1.72}^{+1.54} \times 10^{1}$\\
4 & 1.0 &  800 & $3.20 \times 10^{3}$ & $5.06 \times 10^{-3}$ & $0.58_{-0.06}^{+0.04}$ & $3.61_{-0.70}^{+0.54} \times 10^{1}$\\
4 & 1.0 &  900 & $3.60 \times 10^{3}$ & $3.76 \times 10^{-3}$ & $0.51_{-0.05}^{+0.04}$ & $5.99_{-1.14}^{+0.90} \times 10^{1}$\\
4 & 1.0 &  998 & $3.99 \times 10^{3}$ & $2.91 \times 10^{-3}$ & $0.44_{-0.05}^{+0.04}$ & $9.74_{-2.18}^{+1.73} \times 10^{1}$\\
4 & 1.0 & 1099 & $4.40 \times 10^{3}$ & $2.29 \times 10^{-3}$ & $0.40_{-0.05}^{+0.03}$ & $1.45_{-0.31}^{+0.26} \times 10^{2}$\\
4 & 1.0 & 1200 & $4.80 \times 10^{3}$ & $1.84 \times 10^{-3}$ & $0.39_{-0.05}^{+0.04}$ & $1.83_{-0.44}^{+0.35} \times 10^{2}$\\
4 & 1.0 & 1301 & $5.20 \times 10^{3}$ & $1.50 \times 10^{-3}$ & $0.33_{-0.04}^{+0.03}$ & $2.89_{-0.71}^{+0.57} \times 10^{2}$\\
4 & 1.0 & 1402 & $5.61 \times 10^{3}$ & $1.25 \times 10^{-3}$ & $0.37_{-0.04}^{+0.03}$ & $2.70_{-0.53}^{+0.40} \times 10^{2}$\\
4 & 1.0 & 1497 & $5.99 \times 10^{3}$ & $1.06 \times 10^{-3}$ & $0.31_{-0.05}^{+0.03}$ & $4.42_{-1.25}^{+1.03} \times 10^{2}$\\
4 & 1.0 & 1598 & $6.39 \times 10^{3}$ & $8.97 \times 10^{-4}$ & $0.31_{-0.03}^{+0.02}$ & $4.97_{-0.87}^{+0.68} \times 10^{2}$\\
4 & 1.0 & 1701 & $6.80 \times 10^{3}$ & $7.67 \times 10^{-4}$ & $0.35_{-0.03}^{+0.02}$ & $4.54_{-0.78}^{+0.58} \times 10^{2}$\\
4 & 1.0 & 1803 & $7.21 \times 10^{3}$ & $6.64 \times 10^{-4}$ & $0.35_{-0.03}^{+0.02}$ & $5.10_{-0.87}^{+0.65} \times 10^{2}$\\
4 & 1.0 & 1900 & $7.60 \times 10^{3}$ & $5.82 \times 10^{-4}$ & $0.32_{-0.03}^{+0.02}$ & $6.56_{-0.98}^{+0.66} \times 10^{2}$\\
4 & 1.0 & 1976 & $7.90 \times 10^{3}$ & $5.28 \times 10^{-4}$ & $0.37_{-0.04}^{+0.02}$ & $5.35_{-1.12}^{+0.68} \times 10^{2}$\\
6 & 1.0 &  599 & $5.39 \times 10^{3}$ & $6.95 \times 10^{-3}$ & $0.79_{-0.08}^{+0.06}$ & $1.11_{-0.21}^{+0.17} \times 10^{1}$\\
6 & 1.0 &  701 & $6.31 \times 10^{3}$ & $4.69 \times 10^{-3}$ & $0.58_{-0.08}^{+0.06}$ & $2.83_{-0.73}^{+0.60} \times 10^{1}$\\
6 & 1.0 &  800 & $7.20 \times 10^{3}$ & $3.37 \times 10^{-3}$ & $0.43_{-0.08}^{+0.06}$ & $6.53_{-2.07}^{+1.81} \times 10^{1}$\\
6 & 1.0 &  899 & $8.09 \times 10^{3}$ & $2.52 \times 10^{-3}$ & $0.38_{-0.08}^{+0.06}$ & $1.06_{-0.41}^{+0.37} \times 10^{2}$\\
6 & 1.0 & 1000 & $9.00 \times 10^{3}$ & $1.93 \times 10^{-3}$ & $0.58_{-0.06}^{+0.04}$ & $5.76_{-1.14}^{+0.89} \times 10^{1}$\\
6 & 1.0 & 1100 & $9.90 \times 10^{3}$ & $1.52 \times 10^{-3}$ & $0.54_{-0.07}^{+0.05}$ & $7.98_{-2.02}^{+1.64} \times 10^{1}$\\
6 & 1.0 & 1199 & $1.08 \times 10^{4}$ & $1.23 \times 10^{-3}$ & $0.50_{-0.06}^{+0.04}$ & $1.08_{-0.24}^{+0.20} \times 10^{2}$\\
6 & 1.0 & 1301 & $1.17 \times 10^{4}$ & $9.99 \times 10^{-4}$ & $0.46_{-0.07}^{+0.05}$ & $1.52_{-0.41}^{+0.34} \times 10^{2}$\\
6 & 1.0 & 1400 & $1.26 \times 10^{4}$ & $8.33 \times 10^{-4}$ & $0.47_{-0.06}^{+0.04}$ & $1.72_{-0.41}^{+0.34} \times 10^{2}$\\
6 & 1.0 & 1499 & $1.35 \times 10^{4}$ & $7.01 \times 10^{-4}$ & $0.46_{-0.05}^{+0.03}$ & $2.04_{-0.39}^{+0.29} \times 10^{2}$\\
6 & 1.0 & 1599 & $1.44 \times 10^{4}$ & $5.97 \times 10^{-4}$ & $0.35_{-0.04}^{+0.03}$ & $3.95_{-0.92}^{+0.71} \times 10^{2}$\\
6 & 1.0 & 1697 & $1.53 \times 10^{4}$ & $5.15 \times 10^{-4}$ & $0.36_{-0.04}^{+0.02}$ & $4.16_{-0.78}^{+0.58} \times 10^{2}$\\
6 & 1.0 & 1801 & $1.62 \times 10^{4}$ & $4.44 \times 10^{-4}$ & $0.40_{-0.04}^{+0.03}$ & $3.87_{-0.72}^{+0.53} \times 10^{2}$\\
6 & 1.0 & 1899 & $1.71 \times 10^{4}$ & $3.88 \times 10^{-4}$ & $0.38_{-0.03}^{+0.02}$ & $4.80_{-0.81}^{+0.61} \times 10^{2}$\\
6 & 1.0 & 1977 & $1.78 \times 10^{4}$ & $3.52 \times 10^{-4}$ & $0.37_{-0.06}^{+0.04}$ & $5.46_{-1.74}^{+1.25} \times 10^{2}$\\
8 & 1.0 &  600 & $9.61 \times 10^{3}$ & $5.18 \times 10^{-3}$ & $0.87_{-0.11}^{+0.08}$ & $9.09_{-2.06}^{+1.67} \times 10^{0}$\\
8 & 1.0 &  700 & $1.12 \times 10^{4}$ & $3.54 \times 10^{-3}$ & $0.79_{-0.11}^{+0.08}$ & $1.49_{-0.39}^{+0.33} \times 10^{1}$\\
8 & 1.0 &  800 & $1.28 \times 10^{4}$ & $2.53 \times 10^{-3}$ & $0.73_{-0.08}^{+0.06}$ & $2.29_{-0.48}^{+0.38} \times 10^{1}$\\
8 & 1.0 &  900 & $1.44 \times 10^{4}$ & $1.88 \times 10^{-3}$ & $0.62_{-0.08}^{+0.06}$ & $4.08_{-1.02}^{+0.83} \times 10^{1}$\\
8 & 1.0 &  998 & $1.60 \times 10^{4}$ & $1.45 \times 10^{-3}$ & $0.45_{-0.08}^{+0.06}$ & $9.39_{-2.98}^{+2.57} \times 10^{1}$\\
8 & 1.0 & 1102 & $1.76 \times 10^{4}$ & $1.14 \times 10^{-3}$ & $0.43_{-0.07}^{+0.05}$ & $1.24_{-0.37}^{+0.32} \times 10^{2}$\\
8 & 1.0 & 1199 & $1.92 \times 10^{4}$ & $9.19 \times 10^{-4}$ & $0.49_{-0.05}^{+0.03}$ & $1.14_{-0.22}^{+0.17} \times 10^{2}$\\
8 & 1.0 & 1298 & $2.08 \times 10^{4}$ & $7.54 \times 10^{-4}$ & $0.50_{-0.05}^{+0.04}$ & $1.30_{-0.27}^{+0.21} \times 10^{2}$\\
8 & 1.0 & 1400 & $2.24 \times 10^{4}$ & $6.24 \times 10^{-4}$ & $0.32_{-0.04}^{+0.03}$ & $3.62_{-0.91}^{+0.76} \times 10^{2}$\\
8 & 1.0 & 1499 & $2.40 \times 10^{4}$ & $5.26 \times 10^{-4}$ & $0.31_{-0.05}^{+0.03}$ & $4.52_{-1.23}^{+1.00} \times 10^{2}$\\
8 & 1.0 & 1601 & $2.56 \times 10^{4}$ & $4.46 \times 10^{-4}$ & $0.34_{-0.04}^{+0.03}$ & $4.37_{-0.94}^{+0.72} \times 10^{2}$\\
8 & 1.0 & 1700 & $2.72 \times 10^{4}$ & $3.84 \times 10^{-4}$ & $0.36_{-0.03}^{+0.02}$ & $4.36_{-0.78}^{+0.60} \times 10^{2}$\\
8 & 1.0 & 1800 & $2.88 \times 10^{4}$ & $3.33 \times 10^{-4}$ & $0.42_{-0.03}^{+0.02}$ & $3.58_{-0.53}^{+0.37} \times 10^{2}$\\
8 & 1.0 & 1897 & $3.04 \times 10^{4}$ & $2.92 \times 10^{-4}$ & $0.33_{-0.03}^{+0.02}$ & $6.51_{-1.20}^{+0.91} \times 10^{2}$\\
8 & 1.0 & 1974 & $3.16 \times 10^{4}$ & $2.64 \times 10^{-4}$ & $0.35_{-0.05}^{+0.02}$ & $6.23_{-1.69}^{+0.87} \times 10^{2}$\\

  }{
   2 & 0.0 &  601 & $6.01 \times 10^{2}$ & $2.07 \times 10^{-2}$ & $0.086_{-0.012}^{+0.009}$ & $9.30_{-2.32}^{+1.96} \times 10^{2}$\\
2 & 0.0 &  700 & $7.00 \times 10^{2}$ & $1.41 \times 10^{-2}$ & $0.053_{-0.012}^{+0.010}$ & $3.32_{-1.37}^{+1.29} \times 10^{3}$\\
2 & 0.0 &  800 & $8.00 \times 10^{2}$ & $1.01 \times 10^{-2}$ & $0.061_{-0.010}^{+0.008}$ & $3.32_{-1.03}^{+0.89} \times 10^{3}$\\
2 & 0.0 &  898 & $8.98 \times 10^{2}$ & $7.57 \times 10^{-3}$ & $0.065_{-0.008}^{+0.006}$ & $3.66_{-0.81}^{+0.65} \times 10^{3}$\\
2 & 0.0 & 1000 & $1.00 \times 10^{3}$ & $5.80 \times 10^{-3}$ & $0.069_{-0.008}^{+0.006}$ & $4.07_{-0.89}^{+0.73} \times 10^{3}$\\
\dots & \dots & \dots & \dots & \dots & \dots & \dots\\

  }
 \enddata
 \iftoggle{fulltables}{}{
  \tablecomments{Full table available as online machine-readable version.}
 }
\end{deluxetable*}

\iftoggle{fulltables}{
 \clearpage
 \startlongtable
}{}
\begin{deluxetable*}{crccll}
 \tablecaption{
  \label{tbl:ULs_n5_freqhough}
  \freqhough search sensitivities
  estimated from simulated signals (injections)
  following the power-law spindown model
  with braking index \mbox{$n=5$}.
  Each row corresponds to injections marginalized
  over a 50\,Hz band in $\fref$
  (where \mbox{$\fref=\fgw(\tref=\tc+\toffset)$})
  and random $\cos\iota$;
  sensitivities are at \ULperc confidence.
 }
 \tablehead{
  \colhead{$\Tsft$\,[s]} &
  \colhead{$\fref$\,[Hz]} &
  \colhead{$\tau$\, [s]} &
  \colhead{$\epsilon$} &
  \colhead{$\dUL$\,[Mpc]} &
  \colhead{$\EUL\,[\Msun c^2]$}
 }
 \startdata
  \iftoggle{fulltables}{
   2 &  390 & $4.15 \times 10^{2}$ & $5.91 \times 10^{-2}$ & $0.78_{-0.04}^{+0.05}$ & $4.76_{-0.50}^{+0.59} \times 10^{0}$\\
2 &  440 & $4.65 \times 10^{2}$ & $4.39 \times 10^{-2}$ & $0.83_{-0.05}^{+0.06}$ & $5.42_{-0.62}^{+0.75} \times 10^{0}$\\
2 &  490 & $5.15 \times 10^{2}$ & $3.36 \times 10^{-2}$ & $0.79_{-0.05}^{+0.06}$ & $7.39_{-0.89}^{+1.08} \times 10^{0}$\\
2 &  540 & $5.65 \times 10^{2}$ & $2.64 \times 10^{-2}$ & $0.72_{-0.04}^{+0.05}$ & $1.06_{-0.12}^{+0.15} \times 10^{1}$\\
2 &  590 & $6.15 \times 10^{2}$ & $2.12 \times 10^{-2}$ & $0.75_{-0.05}^{+0.06}$ & $1.19_{-0.16}^{+0.20} \times 10^{1}$\\
2 &  640 & $6.65 \times 10^{2}$ & $1.73 \times 10^{-2}$ & $0.67_{-0.04}^{+0.05}$ & $1.75_{-0.23}^{+0.28} \times 10^{1}$\\
2 &  690 & $7.15 \times 10^{2}$ & $1.44 \times 10^{-2}$ & $0.57_{-0.04}^{+0.04}$ & $2.82_{-0.34}^{+0.42} \times 10^{1}$\\
2 &  740 & $7.65 \times 10^{2}$ & $1.21 \times 10^{-2}$ & $0.63_{-0.04}^{+0.05}$ & $2.65_{-0.36}^{+0.45} \times 10^{1}$\\
2 &  790 & $8.15 \times 10^{2}$ & $1.03 \times 10^{-2}$ & $0.52_{-0.03}^{+0.04}$ & $4.39_{-0.52}^{+0.64} \times 10^{1}$\\
2 &  840 & $8.65 \times 10^{2}$ & $8.83 \times 10^{-3}$ & $0.55_{-0.04}^{+0.04}$ & $4.54_{-0.59}^{+0.73} \times 10^{1}$\\
2 &  890 & $9.15 \times 10^{2}$ & $7.64 \times 10^{-3}$ & $0.60_{-0.05}^{+0.06}$ & $4.23_{-0.64}^{+0.82} \times 10^{1}$\\
2 &  940 & $9.65 \times 10^{2}$ & $6.67 \times 10^{-3}$ & $0.49_{-0.03}^{+0.04}$ & $7.04_{-0.92}^{+1.14} \times 10^{1}$\\
2 &  990 & $1.02 \times 10^{3}$ & $5.87 \times 10^{-3}$ & $0.51_{-0.04}^{+0.05}$ & $7.15_{-1.04}^{+1.33} \times 10^{1}$\\
2 & 1090 & $1.12 \times 10^{3}$ & $4.62 \times 10^{-3}$ & $0.40_{-0.03}^{+0.03}$ & $1.39_{-0.18}^{+0.22} \times 10^{2}$\\
2 & 1140 & $1.16 \times 10^{3}$ & $4.13 \times 10^{-3}$ & $0.38_{-0.02}^{+0.03}$ & $1.75_{-0.22}^{+0.27} \times 10^{2}$\\
2 & 1190 & $1.22 \times 10^{3}$ & $3.71 \times 10^{-3}$ & $0.34_{-0.02}^{+0.02}$ & $2.38_{-0.28}^{+0.35} \times 10^{2}$\\
2 & 1240 & $1.26 \times 10^{3}$ & $3.35 \times 10^{-3}$ & $0.34_{-0.02}^{+0.03}$ & $2.50_{-0.31}^{+0.39} \times 10^{2}$\\
2 & 1290 & $1.32 \times 10^{3}$ & $3.04 \times 10^{-3}$ & $0.45_{-0.04}^{+0.05}$ & $1.60_{-0.27}^{+0.35} \times 10^{2}$\\
2 & 1340 & $1.36 \times 10^{3}$ & $2.76 \times 10^{-3}$ & $0.32_{-0.02}^{+0.02}$ & $3.40_{-0.43}^{+0.53} \times 10^{2}$\\
2 & 1390 & $1.42 \times 10^{3}$ & $2.52 \times 10^{-3}$ & $0.34_{-0.02}^{+0.03}$ & $3.29_{-0.46}^{+0.58} \times 10^{2}$\\
2 & 1440 & $1.46 \times 10^{3}$ & $2.31 \times 10^{-3}$ & $0.33_{-0.02}^{+0.03}$ & $3.66_{-0.51}^{+0.65} \times 10^{2}$\\
2 & 1490 & $1.52 \times 10^{3}$ & $2.12 \times 10^{-3}$ & $0.31_{-0.02}^{+0.03}$ & $4.45_{-0.62}^{+0.79} \times 10^{2}$\\
2 & 1540 & $1.56 \times 10^{3}$ & $1.95 \times 10^{-3}$ & $0.29_{-0.02}^{+0.02}$ & $5.26_{-0.72}^{+0.90} \times 10^{2}$\\
2 & 1590 & $1.62 \times 10^{3}$ & $1.80 \times 10^{-3}$ & $0.30_{-0.02}^{+0.03}$ & $5.55_{-0.80}^{+1.02} \times 10^{2}$\\
2 & 1640 & $1.66 \times 10^{3}$ & $1.67 \times 10^{-3}$ & $0.32_{-0.03}^{+0.03}$ & $5.11_{-0.80}^{+1.05} \times 10^{2}$\\
2 & 1690 & $1.72 \times 10^{3}$ & $1.55 \times 10^{-3}$ & $0.27_{-0.02}^{+0.02}$ & $7.25_{-1.05}^{+1.33} \times 10^{2}$\\
2 & 1740 & $1.76 \times 10^{3}$ & $1.44 \times 10^{-3}$ & $0.24_{-0.02}^{+0.02}$ & $9.92_{-1.31}^{+1.63} \times 10^{2}$\\
2 & 1790 & $1.82 \times 10^{3}$ & $1.34 \times 10^{-3}$ & $0.29_{-0.02}^{+0.03}$ & $7.35_{-1.18}^{+1.55} \times 10^{2}$\\
2 & 1890 & $1.92 \times 10^{3}$ & $1.17 \times 10^{-3}$ & $0.29_{-0.03}^{+0.03}$ & $8.24_{-1.40}^{+1.88} \times 10^{2}$\\
2 & 1940 & $1.96 \times 10^{3}$ & $1.10 \times 10^{-3}$ & $0.28_{-0.03}^{+0.03}$ & $9.20_{-1.61}^{+2.18} \times 10^{2}$\\
2 & 1990 & $2.02 \times 10^{3}$ & $1.03 \times 10^{-3}$ & $0.28_{-0.03}^{+0.03}$ & $9.83_{-1.70}^{+2.30} \times 10^{2}$\\
4 &  290 & $1.26 \times 10^{3}$ & $6.14 \times 10^{-2}$ & $0.97_{-0.07}^{+0.08}$ & $1.71_{-0.23}^{+0.29} \times 10^{0}$\\
4 &  340 & $1.46 \times 10^{3}$ & $4.15 \times 10^{-2}$ & $0.98_{-0.07}^{+0.09}$ & $2.31_{-0.33}^{+0.43} \times 10^{0}$\\
4 &  390 & $1.66 \times 10^{3}$ & $2.96 \times 10^{-2}$ & $0.72_{-0.04}^{+0.05}$ & $5.64_{-0.66}^{+0.81} \times 10^{0}$\\
4 &  440 & $1.86 \times 10^{3}$ & $2.19 \times 10^{-2}$ & $0.74_{-0.05}^{+0.06}$ & $6.83_{-0.89}^{+1.10} \times 10^{0}$\\
4 &  490 & $2.06 \times 10^{3}$ & $1.68 \times 10^{-2}$ & $0.71_{-0.05}^{+0.06}$ & $9.15_{-1.25}^{+1.57} \times 10^{0}$\\
4 &  540 & $2.26 \times 10^{3}$ & $1.32 \times 10^{-2}$ & $0.73_{-0.06}^{+0.07}$ & $1.06_{-0.16}^{+0.20} \times 10^{1}$\\
4 &  590 & $2.46 \times 10^{3}$ & $1.06 \times 10^{-2}$ & $0.61_{-0.04}^{+0.05}$ & $1.79_{-0.25}^{+0.32} \times 10^{1}$\\
4 &  640 & $2.66 \times 10^{3}$ & $8.67 \times 10^{-3}$ & $0.54_{-0.04}^{+0.04}$ & $2.65_{-0.34}^{+0.43} \times 10^{1}$\\
4 &  690 & $2.86 \times 10^{3}$ & $7.19 \times 10^{-3}$ & $0.73_{-0.07}^{+0.08}$ & $1.72_{-0.30}^{+0.41} \times 10^{1}$\\
4 &  740 & $3.06 \times 10^{3}$ & $6.05 \times 10^{-3}$ & $0.50_{-0.03}^{+0.04}$ & $4.28_{-0.56}^{+0.70} \times 10^{1}$\\
4 &  790 & $3.26 \times 10^{3}$ & $5.14 \times 10^{-3}$ & $0.46_{-0.03}^{+0.04}$ & $5.59_{-0.72}^{+0.90} \times 10^{1}$\\
4 &  840 & $3.46 \times 10^{3}$ & $4.41 \times 10^{-3}$ & $0.41_{-0.03}^{+0.03}$ & $8.02_{-0.98}^{+1.19} \times 10^{1}$\\
4 &  890 & $3.66 \times 10^{3}$ & $3.82 \times 10^{-3}$ & $0.50_{-0.04}^{+0.05}$ & $6.06_{-0.94}^{+1.22} \times 10^{1}$\\
4 &  940 & $3.86 \times 10^{3}$ & $3.34 \times 10^{-3}$ & $0.40_{-0.03}^{+0.03}$ & $1.08_{-0.14}^{+0.18} \times 10^{2}$\\
4 &  990 & $4.06 \times 10^{3}$ & $2.93 \times 10^{-3}$ & $0.30_{-0.02}^{+0.02}$ & $2.13_{-0.23}^{+0.27} \times 10^{2}$\\
4 & 1040 & $4.26 \times 10^{3}$ & $2.59 \times 10^{-3}$ & $0.27_{-0.01}^{+0.02}$ & $2.89_{-0.30}^{+0.35} \times 10^{2}$\\
4 & 1090 & $4.46 \times 10^{3}$ & $2.31 \times 10^{-3}$ & $0.29_{-0.02}^{+0.02}$ & $2.79_{-0.32}^{+0.38} \times 10^{2}$\\
4 & 1140 & $4.66 \times 10^{3}$ & $2.06 \times 10^{-3}$ & $0.28_{-0.02}^{+0.02}$ & $3.14_{-0.36}^{+0.44} \times 10^{2}$\\
4 & 1190 & $4.86 \times 10^{3}$ & $1.86 \times 10^{-3}$ & $0.27_{-0.02}^{+0.02}$ & $3.75_{-0.44}^{+0.53} \times 10^{2}$\\
4 & 1240 & $5.06 \times 10^{3}$ & $1.67 \times 10^{-3}$ & $0.25_{-0.01}^{+0.02}$ & $4.88_{-0.56}^{+0.67} \times 10^{2}$\\
4 & 1290 & $5.26 \times 10^{3}$ & $1.52 \times 10^{-3}$ & $0.23_{-0.01}^{+0.02}$ & $5.81_{-0.66}^{+0.79} \times 10^{2}$\\
4 & 1340 & $5.46 \times 10^{3}$ & $1.38 \times 10^{-3}$ & $0.25_{-0.02}^{+0.02}$ & $5.63_{-0.70}^{+0.86} \times 10^{2}$\\
4 & 1390 & $5.66 \times 10^{3}$ & $1.26 \times 10^{-3}$ & $0.24_{-0.02}^{+0.02}$ & $6.19_{-0.79}^{+0.97} \times 10^{2}$\\
4 & 1440 & $5.86 \times 10^{3}$ & $1.15 \times 10^{-3}$ & $0.27_{-0.02}^{+0.02}$ & $5.39_{-0.79}^{+1.01} \times 10^{2}$\\
4 & 1490 & $6.06 \times 10^{3}$ & $1.06 \times 10^{-3}$ & $0.18_{-0.01}^{+0.01}$ & $1.30_{-0.14}^{+0.16} \times 10^{3}$\\
8 &  290 & $5.04 \times 10^{3}$ & $3.07 \times 10^{-2}$ & $0.99_{-0.08}^{+0.10}$ & $1.65_{-0.27}^{+0.35} \times 10^{0}$\\
8 &  340 & $5.84 \times 10^{3}$ & $2.07 \times 10^{-2}$ & $0.93_{-0.08}^{+0.09}$ & $2.55_{-0.41}^{+0.54} \times 10^{0}$\\
8 &  390 & $6.64 \times 10^{3}$ & $1.48 \times 10^{-2}$ & $0.77_{-0.06}^{+0.07}$ & $4.88_{-0.72}^{+0.93} \times 10^{0}$\\
8 &  440 & $7.44 \times 10^{3}$ & $1.10 \times 10^{-2}$ & $0.89_{-0.08}^{+0.10}$ & $4.66_{-0.84}^{+1.14} \times 10^{0}$\\
8 &  490 & $8.24 \times 10^{3}$ & $8.40 \times 10^{-3}$ & $0.53_{-0.03}^{+0.04}$ & $1.61_{-0.20}^{+0.25} \times 10^{1}$\\
8 &  540 & $9.04 \times 10^{3}$ & $6.61 \times 10^{-3}$ & $0.76_{-0.07}^{+0.09}$ & $9.79_{-1.76}^{+2.41} \times 10^{0}$\\
8 &  590 & $9.84 \times 10^{3}$ & $5.30 \times 10^{-3}$ & $0.74_{-0.07}^{+0.09}$ & $1.23_{-0.23}^{+0.33} \times 10^{1}$\\
8 &  640 & $1.06 \times 10^{4}$ & $4.34 \times 10^{-3}$ & $0.66_{-0.06}^{+0.08}$ & $1.81_{-0.32}^{+0.44} \times 10^{1}$\\
8 &  690 & $1.14 \times 10^{4}$ & $3.60 \times 10^{-3}$ & $0.66_{-0.06}^{+0.08}$ & $2.12_{-0.39}^{+0.54} \times 10^{1}$\\
8 &  740 & $1.22 \times 10^{4}$ & $3.02 \times 10^{-3}$ & $0.62_{-0.06}^{+0.08}$ & $2.71_{-0.51}^{+0.70} \times 10^{1}$\\
8 &  790 & $1.30 \times 10^{4}$ & $2.57 \times 10^{-3}$ & $0.54_{-0.05}^{+0.06}$ & $4.08_{-0.71}^{+0.96} \times 10^{1}$\\
8 &  840 & $1.38 \times 10^{4}$ & $2.21 \times 10^{-3}$ & $0.74_{-0.09}^{+0.13}$ & $2.45_{-0.58}^{+0.90} \times 10^{1}$\\
8 &  890 & $1.46 \times 10^{4}$ & $1.91 \times 10^{-3}$ & $0.55_{-0.06}^{+0.07}$ & $4.99_{-0.97}^{+1.37} \times 10^{1}$\\
8 &  940 & $1.54 \times 10^{4}$ & $1.67 \times 10^{-3}$ & $0.65_{-0.08}^{+0.11}$ & $4.01_{-0.94}^{+1.45} \times 10^{1}$\\
8 &  990 & $1.62 \times 10^{4}$ & $1.47 \times 10^{-3}$ & $0.58_{-0.07}^{+0.09}$ & $5.58_{-1.27}^{+1.92} \times 10^{1}$\\
8 & 1040 & $1.70 \times 10^{4}$ & $1.30 \times 10^{-3}$ & $0.45_{-0.05}^{+0.06}$ & $1.03_{-0.20}^{+0.28} \times 10^{2}$\\
8 & 1090 & $1.78 \times 10^{4}$ & $1.15 \times 10^{-3}$ & $0.45_{-0.05}^{+0.06}$ & $1.11_{-0.23}^{+0.32} \times 10^{2}$\\
8 & 1140 & $1.86 \times 10^{4}$ & $1.03 \times 10^{-3}$ & $0.46_{-0.05}^{+0.07}$ & $1.18_{-0.25}^{+0.37} \times 10^{2}$\\
8 & 1190 & $1.94 \times 10^{4}$ & $9.28 \times 10^{-4}$ & $0.34_{-0.03}^{+0.04}$ & $2.30_{-0.40}^{+0.55} \times 10^{2}$\\

  }{
   2 &  390 & $4.15 \times 10^{2}$ & $5.91 \times 10^{-2}$ & $0.78_{-0.04}^{+0.05}$ & $4.76_{-0.50}^{+0.59} \times 10^{0}$\\
2 &  440 & $4.65 \times 10^{2}$ & $4.39 \times 10^{-2}$ & $0.83_{-0.05}^{+0.06}$ & $5.42_{-0.62}^{+0.75} \times 10^{0}$\\
2 &  490 & $5.15 \times 10^{2}$ & $3.36 \times 10^{-2}$ & $0.79_{-0.05}^{+0.06}$ & $7.39_{-0.89}^{+1.08} \times 10^{0}$\\
2 &  540 & $5.65 \times 10^{2}$ & $2.64 \times 10^{-2}$ & $0.72_{-0.04}^{+0.05}$ & $1.06_{-0.12}^{+0.15} \times 10^{1}$\\
2 &  590 & $6.15 \times 10^{2}$ & $2.12 \times 10^{-2}$ & $0.75_{-0.05}^{+0.06}$ & $1.19_{-0.16}^{+0.20} \times 10^{1}$\\
\dots & \dots & \dots & \dots & \dots & \dots\\

  }
 \enddata
 \iftoggle{fulltables}{}{
  \tablecomments{Full table available as online machine-readable version.}
 }
\end{deluxetable*}

\iftoggle{fulltables}{
 \clearpage
}{}

\bibliography{../pmlong.bib}

\iftoggle{endauthorlist}{
  %
  % Restore the author, affiliation and maketitle commands.
  %
  \let\author\myauthor
  \let\affiliation\myaffiliation
  \let\maketitle\mymaketitle
  \title{Authors}
  \iftoggle{fullauthorlist}{
   %\input{LSC-Virgo-Authors-Feb-2018-aas.tex}
    % this version has opt-ins/outs
  }{
   \author{B.~P.~Abbott}
   \author{others}
  }
  \maketitle
}

\end{document}